# Warm Gas in the Inner Disks around Young Intermediate Mass Stars

Sean D. Brittain[1]
*Department of Physics & Astronomy, Clemson University, Clemson, SC 29631*
sbritt@clemson.edu

Theodore Simon
*Institute for Astronomy, University of Hawaii, Honolulu, HI 96822*

Joan R. Najita
*National Optical Astronomy Observatory, 950 N. Cherry Ave, Tucson, AZ 85719*

Terrence W. Rettig
*Center for Astrophysics, University of Notre Dame, Notre Dame, IN 46556*

**Abstract**
The characterization of gas in the inner disks around young stars is of particular interest because of its connection to planet formation. In order to study the gas in inner disks, we have obtained high-resolution K-band and M-band spectroscopy of 14 intermediate mass young stars. In sources that have optically thick inner disks, i.e. E(K-L)>1, our detection rate of the ro-vibrational CO transitions is 100% and the gas is thermally excited. Of the five sources that do not have optically thick inner disks, we only detect the ro-vibrational CO transitions from HD 141569. In this case, we show that the gas is excited by UV fluorescence and that the inner disk is devoid of gas and dust. We discuss the plausibility of the various scenarios for forming this inner hole. Our modeling of the UV fluoresced gas suggests an additional method by which to search for and/or place stringent limits on gas in dust depleted regions in disks around Herbig Ae/Be stars.

---

[1] Michelson Postdoctoral Fellow

## 1. Introduction

A comprehensive picture of star and planet formation requires an understanding of the evolution of dust *and* gas in circumstellar disks. It is through circumstellar disks that stars accrete material, the angular momentum evolution of the star is regulated, and planet formation may occur. While a great deal of progress has been made toward understanding how dust evolves in the inner disk (e.g. Natta et al. 2006 and references therein), the evolution of gas in the inner disk is much less clear. Furthermore, it is not clear whether or not gas and dust dissipate on the same timescale (Najita et al. 2006a). Indeed the basic question of how long primordial gas survives in the disk remains unresolved.

A number of approaches have been taken to investigate the gas in disks around young stars. Studies of cold gas in the outer disk (typically at distances greater than 100 AU from the star) rely on (sub)mm observations of pure rotational transitions of such molecules as CO, $HCO^+$ and HCN (e.g. Dent et al. 2005; Zuckerman, Forveille & Kastner 1995; Beckwith & Sargent 1991). At the other extreme, measurements of accreting gas onto the star rely upon optical observations of such atomic species as sodium and hydrogen (e.g. Muzerolle et al. 1998, 2001).

The measurement of gas in the disk at distances of 0.1-50AU from the star is of particular interest due to its bearing on the potential for planet formation. In principle the most direct way to probe this gas is by the observation of the most abundant molecule, $H_2$. However, thermally emitting $H_2$ gas has relatively steep excitation requirements (the first excited energy level lies 500 K above the ground state) as well as exceedingly small transition probabilities. Because of these characteristics, a vast reservoir of $H_2$ can remain hidden in circumstellar disks at the relatively cool temperatures of ~100 K thought to characterize the giant planet formation region of the disk. The electronic and ro-vibrational transitions of $H_2$, which can be excited *non-thermally*, have proven to be successful probes of disks. Using ro-vibrational transitions of $H_2$ excited by X-rays, Bary, Weintraub & Kastner (2002) demonstrated the presence of $H_2$ in the weak lined T Tauri star DoAr 21 suggesting that the absence of strong accretion diagnostics (such as H$\alpha$ emission) does not necessarily imply that gas in the circumstellar disk has dissipated. Electronic transitions of $H_2$ fluoresced by Lyman $\alpha$ have been observed at ultraviolet wavelengths from a number of T Tauri stars (Herczeg et al. 2006; Walter et al. 2003) and demonstrate that there can be molecular gas within a few AU of the central star at temperatures of a few 1000 K. Unfortunately, since only the relatively hot gas in the inner disk can be fluoresced by Ly $\alpha$ photons, it is not a useful probe of cool gas at large radii (R>~10 AU).

A complementary diagnostic of gas in the inner disk are the ro-vibrational transitions of CO. Previous work has shown that the ro-vibrational lines of CO are sensitive probes of circumstellar gas and are well suited to exploring conditions within the inner, planet-forming regions of disks (Najita et al 2000, 2003, 2006a; Brittain et al. 2003; Blake & Boogert 2004; Rettig et al. 2004, 2006). These lines can be excited thermally or by UV fluorescence. Infrared observations can detect an amount of CO much smaller than an Earth mass, well below the threshold necessary to circularize the orbits of terrestrial planets. Although CO is not a useful probe of the total mass of circumstellar gas at large column densities, its presence can be used to trace the "skin" of the gas disk.

An observational clarification of how the gas evolves in the inner disks of young stars is central to an understanding of early planet formation. If planets form by a two-step process where grains first agglomerate into a large rocky core followed by the accretion of a gaseous envelope (e.g. Lissauer 1993), then the disk should go through a gas-rich/dust-poor stage. Alternatively, if planets form by a gravitational instability (e.g. Boss 1998), then the gas and dust should co-evolve. In addition to adjudicating among models of planet formation, observation of gas in the optically thin disks can guide our interpretation of dust gaps and rings observed around young stars because dust-gas interactions in such disks can lead to dust debris morphologies that mimic sculpting by an embedded companion (Takeuchi & Artymowicz 2001).

But is there any reason to suspect that there may be gas in disks that are dust depleted? Clearly there is abundant gas in optically thick accretion disks (Kenyon & Hartmann 1995). However, some systems with optically thin inner disks are also accreting, so they must have abundant gas (e.g., TW Hya). Indeed, in the case of TW Hya, both UV fluoresced $H_2$ (Herczeg et al. 2006) and ro-vibrational CO emission is detected (Rettig et al. 2004; Najita et al. 2006a; Blake, personal comm.) But what about optically thin disks that are not accreting at a measurable rate? One possibility is that they are not accreting because they are gas-poor. An alternative is that they might have low viscosity and be gas-rich; the lack of accretion or dust in the inner region in these disks might reduce the heating of the disk atmosphere making thermal emission from diagnostics such as CO more difficult to detect. Indeed, there is an observed correlation between CO emission strength and accretion rate (or equivalently K-L excess) among T Tauri stars (Najita et al. 2003). Modeling of the heating of circumstellar disks by Glassgold et al. (2004) indicates a physical basis for this correlation as they find that accretion related processes might contribute substantially to heating T Tauri disk atmospheres. Similarly, photoelectric heating of small grains in the disk may also be important. As the number of small grains decreases in optically thin disks (i.e. the K-L excess decreases), then the effectiveness of photoelectric heating will diminish as well. Thus, for either reason, dust-depleted, gas-rich disks may not reveal thermally excited gas emission lines.

Based on the similarity of disks around classical T Tauri stars and HAeBe stars, one might also expect accretion to play a role in the excitation of CO in disks around HAeBe stars. These similarities have motivated Muzerolle et al. (2004) to apply their magnetospheric accretion model to the case of the HAe star UX Ori. They find they can reproduce the Hα and Hβ line profiles with their magnetospheric accretion models. Similarly, van den Ancker (2004) finds that the relationship between the luminosity of the UV excess and the Br γ luminosity of HAeBe stars is the same as that found for T Tauri stars. We will explore the possibility that the Br γ emission provides a reliable measure of the accretion rate in more detail in §5.2.

Another means of exciting CO gas around HAeBe stars is through UV fluorescence (Brittain et al. 2003). The CO molecule has bound excited electronic states whose transitions occur at UV wavelengths. When the excited states are populated, the molecule relaxes into excited vibrational levels in the electronic ground state resulting in bright ro-vibrational emission lines. We present modeling that shows the conditions where we expect UV fluorescence and discuss its utility for measuring trace amounts of molecular gas in dust depleted disks.

After presenting the data (§3.1; §3.2), we address the special case of HD 141569 (§3.2.2), present modeling of UV fluorescence (§4), address the role of accretion and fluorescence in exciting CO emission (§5.1-§5.3), and describe implications of the (non)detections (§5.4). The goal of this paper is to demonstrate the feasibility of using CO to diagnose the presence of gas in disks around HAeBe stars and present a theoretical framework for interpreting the non-detections of CO in optically thin regions of transitional disks.

## 2. Observations

We have observed a heterogeneous sample of 14 young intermediate mass stars (2-10 $M_{\odot}$) to measure the emission from H I (Br γ) and CO (Δv=1) (Table 1). The targets include two intermediate mass T Tauri stars (IMTTS), seven Herbig AeBe (HAeBe) stars, four transitional

HAeBe stars, and one heavily embedded intermediate-mass object that has not been definitively classified. The Br γ and CO spectra are presented in figures 1 and 2. The sources were selected to span a broad range of NIR excess (-0.08<K-L<2.2) and thus a broad range of circumstellar evolution.

The observations were obtained at the W. M. Keck Observatory. The NIRSPEC echelle spectrograph (McLean et al. 1998) at Keck 2 provided a resolving power of $\lambda/\Delta\lambda = 25{,}000$. A series of flats and darks were used to remove systematic effects at each grating setting. The 2-dimensional frames were cleaned of systematically hot and dead pixels as well as cosmic ray hits, and were then resampled spatially and spectrally. This initial processing results in frames for which spectral and spatial dimensions are orthogonal, falling along rows and columns respectively (DiSanti et al. 2001).

A significant thermal (~300 K) continuum background dominates the M-band spectra. Night sky emission lines are superimposed on this background. The intensities of the telluric lines depend not only on the air mass, but also on the column burden of atmospheric water vapor, which can vary both temporally and spatially over the course of the night. In order to cancel most of the atmospheric + background, the telescope was nodded by a small distance, typically 15″, along the slit. The nod pattern was from the position (A) to the position (B) in an {A, B, B, A} sequence, with each step corresponding to 1 minute of integration time. Combining the scans as {A-B-B+A}/2 cancels the background to first order. Subsequently, the full width at half-maximum (FWHM) of the spatial profiles in the "A" and "B" rows are extracted to obtain the spectra for both positions. The atmospheric transmittance function of the combined spectrum was modeled using the Spectrum Synthesis Program (SSP, Kunde & Maguire 1974), which accesses the updated 2000HITRAN molecular database (Rothman 2003). For each grating setting listed in Table 1, the optimized model establishes the column burden of absorbing atmospheric species, the spectral resolving power, and the wavelength calibration. The wavelength coverage of each setting is presented in Table 2.

## 3. Results

### 3.1 HI

The Br γ transition was detected in 12 of the 14 sources (Fig. 1). It is seen in absorption in two of the sources, as a singly peaked emission line in seven of the sources, and as a double-peaked transition in three of the sources. The centroid of the absorption lines and the singly peaked emission lines was determined from a Gaussian fit to the profile whereas the centroid of the double peaked lines was determined from the center of the full width at zero intensity (FWZI).

The FWZI of the single peaked Br γ lines range from 300 to 800 km $s^{-1}$, and the FWHM of the emission features range from 48 to 350 km $s^{-1}$. Clearly, the line profiles are not dominated by thermal broadening. The emission profiles are qualitatively similar to those of cTTSs presented by Folha & Emerson (2001) in that they are broad and do not reveal the blue-shifted absorption component that is often seen in the Hα profiles of HAeBes (cf., for example, the profile of AB Aur in Finkenzeller & Mundt 1984). The absence of such absorption indicates that the single-peaked Br γ transitions do not originate in a wind, which is not surprising since the n=4 energy level of hydrogen is less likely to be populated in the stellar wind.

The emission line from most of the sources is at rest or has a slight blue-shift (typically a few km $s^{-1}$) relative to the star (Table 3), though HD 141569, HD 163296, and HR 5999 have substantially larger blue-shifts (15-30 km $s^{-1}$). The one exception is HD 250550, which has a 6 km $s^{-1}$ red-shift; the origin of the red-shift is not clear. Blue-shifted emission is a characteristic of emission lines formed in accreting gas (Edwards et al. 1994; Najita et al. 1996; Muzerolle et al. 1998). The atomic emission line profiles of HAeBe stars have been extensively studied, though there seems to be no consensus as to their origin. Attempts at an explanation of the profiles have included appeals to winds (e.g. Strafella et al. 1998; Catala 1989), circumstellar gas (Dunkin et al. 1997a), stellar activity (Böhm & Catala 1995) and magnetospheric accretion flows (van den Ancker 2004). In §5.2 we explore the possibility that the Br $\gamma$ emission provides a reliable measure of the accretion rate and examine the relationship between the flux of the CO emission and Br $\gamma$ emission.

The profile of the double peaked lines is similar to those of the Balmer emission lines that are commonly seen in late B/early A shell stars (Andrillat 1983; Andrillat, Jaschek, & Jaschek 1986; Hanuschik et al. 1996a). The double peaked HI lines are formed by the Keplerian orbit of gas in a compact disk formed by the ejection of material from a star rotating near its breakup velocity (Gies et al. 2006, Meilland et al. 2006, Hanuschik 1996b). Two of the sources we observe with double-peaked lines, HD141569 and HD158643, have the largest projected rotational velocities ($v \sin i$ = 256 and 236 km $s^{-1}$ respectively). In these two stars, the wings of a photospheric feature are also visible (compare HD 158643 and HD 141569 in Fig. 1). In order to extract the flux of the Br $\gamma$ emission component of these stars, we fit a synthetic Br $\gamma$ absorption line to the ratioed data. The line profile was calculated using the PHOENIX code (see Aufdenberg, Hauschildt & Baron 1999). The input spectral type, gravity and projected rotational velocity were taken from Dunkin et al. (1997b).

The fit and residuals are plotted in figure 3. The emission flux for each line was determined by summing the observed flux above the model curve. In the case of the double peaked emission line observed from HD 58647, we simply summed the flux of the line since no photospheric absorption was observed. The weak photospheric absorption is likely a result of the large K-band excess ($\Delta$K=1.4 mag) of HD 58647 compared to either HD 158643 ($\Delta$K=0.37) or HD 141569 ($\Delta$K=0.05; magnitudes from Malfait et al. 1998), so the IR excess veils the wings of the Br $\gamma$ absorption line.

HD 141569 and HD 158643 are particularly well-studied HAeBe stars, so there is a wealth of data on these sources. Interestingly, the profile of H$\alpha$ for each source is similar to the profile of Br $\gamma$. For HD 158643 the projected velocity measured from the FWZI and peak separation of the Br $\gamma$ line are 550±50 km $s^{-1}$ and 250±10 km $s^{-1}$ respectively. The FWZI and peak separation of H$\alpha$ are 600±100 km $s^{-1}$ and 168 km $s^{-1}$ (Dunkin et al. 1997a). The FWZI and peak separation of the $\lambda$8542 CaII IR triplet line are ~500 km $s^{-1}$ and ~200 km $s^{-1}$ respectively (Dunkin et al. 1997a). Similarly, for HD 141569, the projected velocities of the FWZI and peak separation of the Br $\gamma$ line are 700 km $s^{-1}$ and 300 km $s^{-1}$ respectively. The FWZI of H$\alpha$ is 700 km $s^{-1}$ and the peak separation is 242 km $s^{-1}$ (Dunkin et al. 1997a; see also Andrillat et al. 1990). Additionally, the OI $\lambda$8446 line has a FWZI of 800 km $s^{-1}$ and a comparable peak separation (250-285 km $s^{-1}$; Andrillat et al. 1990), whereas the [OI] $\lambda$6300 line has a FWZI of 300 km $s^{-1}$ and peak separation of 75 km $s^{-1}$ (Acke et al. 2005). The similarity of the Br $\gamma$, H$\alpha$, and OI $\lambda$8446 profiles indicates a common origin for the lines, and Dunkin et al. (1997a) claim that the double peaked structure of H$\alpha$ implies that it originates in a circumstellar disk. If these lines do originate in the circumstellar disk, then the H$\alpha$ line from HD 141569 forms in a region that extends from the stellar surface to 0.05AU and the [OI]$\lambda$6300 line originates in a region that extends from 0.05AU to 0.8AU.

### 3.2 CO

#### 3.2.1 General Trends

CO emission was detected in 10 of the 14 sources (Fig. 2). Approximating CO as a rigid rotor, one can demonstrate that the flux of a ro-vibrational emission line is,

$$F_J = \Omega I_{em} = \frac{\Omega C_{em}}{Q_r} \tilde{\nu}^4 (J' + J'' + 1) e^{\frac{-hcBJ'(J'+1)}{kT}} \tag{1}$$

and thus,

$$\frac{k}{hcB} \ln\left(\frac{F_J}{\tilde{\nu}^4 (J' + J'' + 1)}\right) = -\frac{1}{T} J'(J' + 1) + \frac{k}{hcB} \ln\left(\frac{\Omega C_{em}}{Q_r}\right). \tag{2}$$

Here, $k$ is the Boltzmann constant, $h$ the Planck constant, $c$ the speed of light, and $B$ a rotational constant. $F_J$ is the measured flux of each line, $J'$ the upper transition state, $J''$ the lower state, $T$ the temperature of the gas, $\tilde{\nu}$ the frequency of the transition in wavenumbers ($cm^{-1}$), $C_{em}$ a spectroscopic constant, $\Omega$ the solid angle subtended by the beam, and $Q_r$ the partition function for a rigid rotor (Herzberg 1950). $J'(J'+1)$ is plotted versus $(k/hcB)\ ln\{F_J/[\tilde{\nu}^4(J'+J''+1)]\}$ in figure 4 so that the rotational temperature is defined by the negative reciprocal of the linear least-squares slope. In nine of these sources, CO transitions spanning J=0 through J=32 were detected indicative of hot (>$10^3$ K) gas. In one of the sources, HD 141569, only low-J lines (J<20) were detected implying that the gas is cool (Fig. 5, 6).

We find that CO is consistently present in all of the sources with an optically thick inner disk (K-L>1). This result is in accord with the data presented by Blake & Boogert (2004), but it stands in marked contrast with the lower resolution observations of Carmona et al. (2005) who do not detect CO emission in any of the 5 HAeBes they observed. The NIR excess of their targets ranged from 1.1≤E(K-L)≤1.4. It is important to note that the width of emission lines can vary dramatically from source to source (Fig. 2), so any upper limit on a non-detection of an emission line must necessarily assume a line width. In our data the line widths span about an order of magnitude. Also the data acquired by Carmona et al. (2005) were taken at lower resolution (~30 km $s^{-1}$), so the emission lines are not resolved or clearly separated from the telluric absorption features. Thus the significance of these non-detections is not clear.

Our results show that the CO emission lines are generally symmetric and centered on the radial velocity of the star, indicating that the CO gas is most likely associated with the star (Table 3). The one exception is HD 163296 for which the CO emission lines are blue-shifted by 15 km $s^{-1}$ relative to the stellar velocity. The reason for the large discrepancy between the published radial velocity of the star and the Doppler shifts of CO and Br γ is not clear.

The FWHM of the lines span a broad range, from such sources as AB Aur, whose lines are unresolved (or perhaps marginally resolved), to such sources as HD 158643, whose lines are highly broadened (FWHM~200 km $s^{-1}$). The continuum S/N is generally not good enough to infer an accurate FWZI, but the projected velocity implied by the FWZI is generally smaller than the stellar vsin(i). Higher S/N observations are necessary to establish whether the molecular component of the disk extends interior to the co-rotation radius of the disk (i.e. the radius at which the disk and star have the same orbital period). Measuring the truncation radius of the gas disk may be an important tool for estimating the magnetic field strength of HAe stars undergoing magnetospheric accretion.

For the sources with K-L>1, the CO spectra are consistent with hot optically thick gas (see also Blake & Boogert 2004; Brittain et al. 2003). However, in the unusual case of HD 141569 the rotational temperature is ~200 K, although the vibrational temperature is ~5000K. This is indicative of UV fluoresced CO. Because UV fluoresced CO is potentially a valuable probe of trace amounts of gas in optically thin disks, we discuss the data from HD 141569 in more detail.

### 3.2.2 The Special Case of HD 141569

Initial observations of the CO emission lines from HD 141569 have been introduced in Brittain & Rettig (2002) and Brittain et al. (2003). Here we present new, higher signal-to-noise data with greater spectral coverage. The spectra are plotted in figure 6, which shows the data and an atmospheric model fit (Kunde & Maguire 1974). In table 4 we give the integrated flux, line position, and Doppler shift for all of the observed CO lines. A cursory look at the data reveals that the $^{12}$CO v=1-0, v=2-1, v=3-2, v=4-3, v=5-4, and v=6-5 lines all have comparable strengths. The heliocentric Doppler shift of the lines is –7 ± 3 km s$^{-1}$ which is consistent with the heliocentric Doppler shift of HD 141569, –6 ± 5 km s$^{-1}$ (Frisch 1987). The spatial profile of the CO emission lines is not extended beyond the spatial profile of the star.

The HWZI of an emission line formed in a disk corresponds to the inner radius of the emitting gas (e.g. Najita et al. 1996). In the case of HD 141569, the emission lines are only marginally resolved. Fig. 7 shows the v=2-1 R9 line, which was chosen because it is not blended with other emission lines or telluric features. The HWZI of the calibration lamp line profile is 22 km s$^{-1}$, and the HWZI of the v=2-1 R9 emission line is ≤26 km s$^{-1}$. Thus the maximum projected velocity of the gas is 14 km s$^{-1}$. HD 141569 is inclined by 51±3°, its position angle is 356 ± 5° (Weinberger et al. 1999) and the stellar mass is $2.00^{+0.06}_{-0.05}M_{\odot}$ (Merín et al. 2004), so the inner radius of the gas is constrained to >6 AU. Since the line is centrally peaked, it is not formed in a narrow ring at 6AU. On the other hand, the gas is not spatially extended in our 0".7 beam. The distance to HD 141569 is 108±6 pc (Merín et al. 2004), so the emission must originate within ~40 AU of the central star. In section 4.3, we will present more detailed spectral synthesis of the emission profile.

The rotational temperature of the $^{12}$CO vibrational levels v=1through v=4 and the $^{13}$CO vibrational levels v=1 and v=2 is consistent with T~200 K (Fig. 5). Because we can calculate the vibrational populations of v=1 through v=6, we can determine the vibrational temperature. The relative vibrational populations are related by

$$N_v = \frac{N_{total}}{Q_v} e^{-v3122/T} \qquad (3)$$

where $Q_v = 1/(1-e^{-3122/T})$ is the partition function, v is the vibrational level, and 3122 K $(= hc\omega_e/k = E_{vib}/k)$ is the vibrational constant for CO. Thus we plot $ln(N_v)$ vs $vhc\omega_e/k$ such that the negative reciprocal of the slope is the vibrational temperature $T_v$ and the y-intercept is $ln(N_{total}/Q_v)$. Curiously, the vibrational temperature of the gas is 5600 ± 800 K (Fig. 8). This emission pattern (a "cool" rotational population and a "hot" vibrational population) is consistent with excitation of CO gas by ultraviolet fluorescence of the 4$^{th}$ positive system at 0.155μm (Krotkov, Scoville, & Wang 1980). The 4$^{th}$ positive system refers to the electronic transitions between the ground electronic state ($X^1\Sigma^+$) and the first excited singlet state ($A^1\Pi$) that occur at ~1600Å. Data from the International Ultraviolet Explorer (IUE) reveal significant UV flux in this wavelength region. In the following section we demonstrate that UV fluorescence of gas in a truncated, dust-depleted disk can account for the strength and vibrational temperature of the observed CO emission.

## 4. Fluorescence modeling

### 4.1 General Considerations

Krotkov et al. (1980) have presented a thorough overview of the IR signature of UV fluorescence. The probability[2] that a molecule will resonantly scatter a photon is given by

$$g_{ij} = \frac{\pi e^2}{mc^2}\lambda^2 f_{ij}(\pi F_\lambda) \text{ photons s}^{-1} \text{ molecule}^{-1} \qquad (4)$$

where $f_{ij}$ is the oscillator strength of the transition, $F_\lambda$ is the stellar flux in units of photons s$^{-1}$ μm$^{-1}$ cm$^{-2}$, $\lambda$ is the wavelength of the transition and $e^2/mc^2$ is the classical electron radius (Barth 1969; Feldman et al. 1976). Here we adopt the conventional nomenclature for labeling the energy levels and transitions of the CO molecule (e.g., Herzberg 1950).

The population of a vibrational level in the $X^1\Sigma^+$ ground electronic state of CO can be fed by spontaneous and stimulated relaxation from an excited vibrational level in the $A^1\Pi$ electronic state ($A_{A\text{-}X}$ and $g_{A\text{-}X}$ respectively), excitation from a lower level within the electronic ground state($g_{X\text{-}X}$), and spontaneous and stimulated relaxation from an excited vibrational state ($A_{X\text{-}X}$ and $g_{X\text{-}X}$). The levels can be emptied by UV and IR fluorescence ($g_{X\text{-}A}$ and $g_{X\text{-}X}$) and spontaneous and stimulated relaxation ($A_{X\text{-}X}$ and $g_{X\text{-}X}$). The Einstein A's for electronic transitions are of order $10^7$ sec$^{-1}$ and for ro-vibrational transitions are of order 10 sec$^{-1}$. Because $g_{X\text{-}A} >> g_{X\text{-}X}$ we ignore infrared fluorescence. Taking the various foregoing processes into account and assuming a condition of statistical equilibrium, we cast the rate equation for the change in the fractional population of a ground electronic state ($n_X^{v=i}$) as follows:

$$\begin{aligned}\frac{dn_X^{v=i}}{dt} &= \sum_j \left(A_{A-X}^{v'=j,v''=i} + g_{A-X}^{v'=j,v''=i}\right) n_A^{v=j} + A_{X-X}^{v'=i+1,v''=i} n_X^{v=i+1} \\ &\quad - n_X^{v=i}\left(\sum_j g_{X-A}^{v''=i,v'=j} - A_{X-X}^{v'=i,v''=i-1}\right) = 0\end{aligned} \qquad (5)$$

Implicit in the above expression is the assumption that the level populations are in steady state, collisional excitation of vibrational levels are unimportant, and only $\Delta v$=1 transitions are important for the ground electronic state. In figure 9, we illustrate the types of transitions we consider using a three level schematic. The rotational levels are thermalized up to J=20 at densities as low as $n(H_2)=10^6$ $cm^{-3}$, thus the rotational temperature reflects the kinetic temperature of the gas (200 K; Figure 5)[3].

---

[2]Do not confuse g with the commonly used designation for the degeneracy of the state (i.e. $g_J$ for a rotational level of CO is 2J+1).

We have calculated the fluorescent excitation of CO at distances of 0.1 AU, 1 AU and 10 AU from an A0 star (Fig. 10). We adopted the UV radiation of a 10,000K blackbody. The band oscillator strengths were calculated from the Einstein A's presented by Beegle et al. (1999),

$$\left(2-\delta_{0\Lambda''}\right)\lambda_{v'v''}f_{v'v''} = 4\pi\varepsilon_0 \frac{m_e c}{8\pi^2 e^2}\left(2-\delta_{0\Lambda'}\right)\lambda^3_{v'v''}A_{v'v''} \tag{6}$$

where $\frac{\left(2-\delta_{0\Lambda''}\right)}{\left(2-\delta_{0\Lambda'}\right)}$ is the statistical weight of a transition and Λ is the electronic angular momentum of the state (Morton et al. 1994). The value of the transition probabilities and emission rate factors for $X^1\Sigma^+$ v=0 through v=10 and $A^1\Pi$ v=0 through v=9 were adopted in our calculations. Equation 5 was solved for 20 vibrational states ($X^1\Sigma^+$ v=0 through v=9 and $A^1\Pi$ v=0 through v=9) with the requirement that $\Sigma n_i=1$ such that there were 21 equations with 21 unknowns. The system of equations was solved by applying *singular value decomposition*, a numerical technique for solving systems of linear equations, with the IDL routines SVDC and SVSOL. The solution to the equations is the fractional population of each vibrational state. Figure 10 shows that at small radii, ~0.1 AU, the vibrational population is similar to the color temperature of the stellar UV field ~10,000K. At larger radii, as the radiation field of the star grows more dilute, the spontaneous decay of the vibrational levels becomes more significant and "cools" the levels more effectively, causing the vibrational temperature to drop.

**4.2 Application to HD 141569**

We have also calculated the fluorescent excitation of CO distributed from 6-40AU specifically for HD 141569, using low resolution IUE data to calculate the UV radiation field near 1550Å. The spectra were de-reddened using the interstellar extinction law given by Cardelli et al. (1989) and E(B-V)=0.16 (Malfait et al. 1998). For this calculation we assume that extinction by dust in the disk is not significant. The fractional vibrational populations of CO observed toward HD 141569 have been normalized to the calculated value of $n_{v=3}$, and plotted over the calculated populations in figure 11. The excellent agreement between the model and the data confirms that the CO emission observed from HD 141569 is due to UV fluorescence. Thus UV fluorescence can be an important excitation mechanism in dust-depleted disks.

With the vibrational populations of the CO bands in hand, we can now calculate the expected line luminosity of the gas under the following conditions. We assume the CO gas is in hydrostatic equilibrium at 200K (the measured rotational temperature of the CO §3.2) and extends from 6 to 40 AU around a $2M_{\odot}$ star. The scale height of the gas is 0.6 AU at a distance of 6 AU and increases to 5.0 AU at a distance of 40 AU. We have computed the luminosity of the excited gas in increments of 0.1AU concentric rings of the disk (smaller annuli did not appreciably change our results). The UV flux illuminates the front surface of the disk and the top of the ring for each subsequent annulus. The disk is physically truncated above one scale height, and we assume the flux is constant over each ring. Within a given ring we step along the line of sight into the inclined disk atmosphere in increments of N(CO)=2 x $10^{12}$ $cm^{-2}$ such that the increase in the optical depth of the strongest fluorescent transition is $\Delta\tau = 0.2$. We assume the rms linewidth, *b*, is 1.3 km $s^{-1}$. This is based on the turbulent line width of CO in the upper disk of HL Tau which we assume here is representative of disk atmospheres (Brittain et al. 2005). The UV flux from the star is attenuated in each step within the ring, the contribution to the emergent UV fluorescence emission is calculated, and the residual UV flux is allowed to excite the next step. For each ring we calculate the fluorescence of the gas for 500 depths into the disk (or $\tau$=100 for the strongest transition), and we ignore dust extinction in our calculation.

---

[3] The CO rotational levels will be thermalized when the collisional excitation rate, $n(H_2)k_{CO-H_2}$ is comparable to the spontaneous emission rate $A_{J'J''}$. The collision rate constant, $k_{CO-H_2}$, for pure rotational transitions in the ground vibrational state is of order $6\times10^{-11}$ $cm^3$ $s^{-1}$ (Schinke et al. 1985), and the spontaneous emission rate for these transitions range from $10^{-8}$ to $10^{-4}$ $s^{-1}$. Thus the critical density to thermalize the rotational levels of CO in the vibrational ground state is only $\sim10^7$ $cm^{-3}$. Collisions with other partners such as H and He may also be important, and these will lower the overall critical density. Since the rotational transitions within a vibrational level are much slower than ro-vibrational transitions, the rotational population of the ground vibrational state will be reflected in the excited vibrational states as well. Thus the rotational temperature reflects the kinetic temperature of the gas.

The flux of the v=2-1 R9 line, integrated from 6 AU to 40AU, is 4.8x$10^{-15}$ erg/s/$cm^2$ under these conditions. The measured flux of the v=2-1 R9 line is 7.3±0.3 x $10^{-15}$ erg/s/$cm^2$ (Table 4). This line was selected for comparison because it is isolated from CO lines from other vibrational levels and telluric absorption. In our calculation, 50% of the flux originates from 6-15AU and 90% of the flux originates from 6-34 AU (Fig. 12). The reason our calculation underestimates the flux could be because the intrinsic line width we assume is too small. An increase in the intrinsic line width to 2 km $s^{-1}$ brings the computed line flux into agreement with our measurement. Such a width is reasonable for mildly supersonic turbulence at 200 K. For example, Goto et al. (2003) report that CO absorption lines in nearby molecular clouds are turbulently broadened by as much as 3.5 km $s^{-1}$.

In addition to the uncertainty in the turbulent broadening of the gas, there is also some uncertainty in the amount of dust extinction in the disk atmosphere. In the preceding calculation, we assumed that extinction of the UV flux by circumstellar dust was negligible. If the composition of the disk is the same as a dense cloud, then the column density of excited CO, N(CO)~$10^{15}$ $cm^{-2}$, corresponds to a visual extinction of 0.01 magnitudes. For a normal interstellar reddening law, this translates to ~0.1 magnitudes of extinction at 1500 Å, which would have a negligible impact on the fluorescent CO flux. However, we are looking at the "surface" of the molecular gas component of the disk, so it is entirely possible that the CO is underabundant due to photodissociation by the star and/or the interstellar radiation field. If the $CO/H_2$ ratio is more comparable to a diffuse cloud (i.e. $10^{-6}$ rather than $10^{-4}$; compare for example van Dishoeck & Black 1986 and Kulesa & Black 2003) and the gas/dust ratio is ~100, this would imply that the dust extinction is ~10 magnitudes. If $CO/H_2$ is less than $10^{-5}$, the dust in the disk atmosphere would also need to be depleted to allow the star to excite a significant column density of CO. This would be reasonable since significant grain growth and settling is indicated for this system. A detailed calculation of the molecular abundance in the disk atmosphere combined with data like those presented here may provide a means to determine the extent of dust settling in the disk atmosphere.

**4.3 Profile modeling**

Having calculated the intensity of the CO emission as a function of radius, we can now construct a synthetic profile of the emission lines radiated by the gas. We assume the gas is in Keplerian orbit and inclined by 51° (Weinberger et al. 2000) and we divide the disk into annuli 1AU wide. For each annulus, we sum the Doppler shifted flux in 1 km $s^{-1}$ velocity bins. The annuli are summed and convolved with the instrumental profile. Our spectral synthesis indicates that the inner edge of the gas disk must be between 6 AU and 15 AU (Fig. 13). Our best fit is achieved with an inner radius of 9 AU. Goto et al. (2006) present spatially resolved spectra of ro-vibrational emission lines from HD 141569, and they find the inner edge of the CO emission is 11±2 AU consistent with what we deduce from line profile fitting for the assumed stellar mass of 2.0 $M_{\odot}$.

Interestingly, Acke et al. (2005) interpret their observations of the [OI] λ6300 as a dissociation product of OH in the circumstellar disk around HAeBe stars. In the case of HD 141569, if we attribute the [OI] λ6300 line broadening to a Doppler shift from Keplerian orbital motion, the deprojected velocity of the FWZI and the peak separation (300 km $s^{-1}$ and 75 km $s^{-1}$ respectively; see §3.1) indicate the emission originates between ~0.05 and 0.8 AU. However, the CO emission observed toward this source originates at a much larger distance. If the line does originate from the photodissociation of OH, the inner disk is curiously rich in OH but depleted in CO. This appears to call into question the suggestion by Acke et al. (2005) that the [OI] λ6300 emission line in HD 14169 originates from the photodissociation of OH.

Our modeling demonstrates that we can account for the observed strength of the CO emission as well as the relative population of the ground electronic state vibration levels as fluorescent emission driven by the stellar UV flux. The lack of fluorescent CO emission from radii less than 6 AU (which would appear as a broad emission component) suggests that there is little CO interior to this radius. Our upper limit on the column density of CO along the line of sight from the star to the edge of the disk at 6 AU is N(CO)<$10^{15}$ $cm^{-2}$ (Figure 13).

Does the inner-radius of the CO correspond to the inner radius of the gas disk? Merín et al. (2004) estimate that the surface density of the disk at 10AU is 3 g $cm^{-2}$ based on their model of the SED and a gas-to-dust ratio of 100. For a 200 K disk in hydrostatic equilibrium, this suggests an average gas density of ~$10^{10}$ $cm^{-3}$. At these densities virtually all of the carbon is in the form of CO (even at densities as low as 500 $cm^{-3}$, only $A_V$~0.8 mag is required to ensure that virtually all of the carbon is bound up in CO according to van Dishoeck & Black 1986). Given the high gas density at 10 AU, it is unlikely that the inner boundary of the CO emission represents a dissociation front, because the dissociation of CO would require a precipitous drop in the gas density. This raises the interesting question as to why there is such a sharp drop off in the surface density of the disk at 6-15AU, a matter that we discuss further in §5.3.

## 5. Discussion

### 5.1 CO Detections and Infrared Excesses

An important issue in circumstellar disk studies is the degree to which gas and dust co-evolve. If the dust close to the star dissipates leaving behind a gas-rich inner disk, one might expect to detect bright CO emission lines from sources that do not reveal a strong NIR excess. To explore the relationship between the luminosity of the P30 lines and infrared excesses we compare the P30 luminosity with E(K-L) (Fig. 14). Just as the CO P30 line traces hot gas (T>$10^3$K), a K-L excess is indicative of hot dust (T~$10^3$ K; Strom et al. 1989; Skrutskie et al. 1990; Hillenbrand et al. 1992; Malfait et al. 1998; Haisch et al. 2001; Dullemond, Dominik, & Natta 2001). For sources with E(K-L)>0.9 mag, we observe the v=1-0 P30 line in 9 out of 9 sources.

For sources with a secure K-L excess less than 0.6 magnitudes, we do not observe the v=1-0 P30 CO line. Does this mean that there is no CO gas present or that the CO remaining in the inner disk is not sufficiently excited? One might infer from our results that ro-vibrational CO emission simply traces the presence of hot dust and thus dust and gas are co-depleted in the inner disk. The photoelectric heating of small grains in the disk can heat the gas, so a reduction of in K-L for optically thin disks can result in a lower CO emissivity. However, the trend between the infrared excess of the source and the luminosity of the CO emission for T Tauri stars is not restricted to optically thin disks (Najita et al. 2003). The relationship between the K-L excess and detection rate of CO could instead highlight the importance of accretion for heating the disk leading to the collisional excitation of CO. In order to explore whether the weakness of the CO emission in the low K-L sources might be a consequence of a low accretion rate, and potentially reduced accretion heating, we attempt in the next section to use our Br γ measurements as a measure of the stellar accretion rate.

**5.2 CO Detections and the Accretion rate of HAeBe stars**

The accretion rate of T Tauri stars has been measured by numerous methods: U-band excess, blue spectra (Gullbring et al. 1998), and emission line diagnostics (Muzerolle et al. 1998). Unfortunately, the most direct measurement of the accretion luminosity (U-band excess measurements) is challenging for HAeBe stars because the contrast between the accretion shock and near-UV flux of the star is small. van den Ancker (2004) has correlated the accretion luminosity inferred from the FUV luminosity of a dozen HAeBe stars with the Br γ luminosity (determined from low resolution spectra) and found the same relation as that given by Muzerolle et al. (1998),

$$\log\left(\frac{L_{acc}}{L_{\odot}}\right) = (1.26 \pm 0.19)\log\left(\frac{L_{Br\gamma}}{L_{\odot}}\right) + (4.43 \pm 0.79) \quad (7)$$

This relationship was calibrated by van den Ancker (2004) down to accretion rates of $\dot{M} \approx 10^{-7} M_{\odot}\, yr^{-1}$. There are a number of caveats to note in such a study. First, the lack of contrast between stellar continuum and accretion shock may lead to underestimate of line flux and thus accretion rate. Second, the convolution of stellar absorption and circumstellar emission at low spectral resolution may lead to an underestimate of the line flux. Third, the orientation of the accretion shock may affect the line luminosity and introduce further scatter to the relationship. Fourth, HI emission can originate in circumstellar environments other than accretion shocks. Finally, the rotational speed of HAeBes makes the co-rotation radius of the disk very small, so that the emitting area of the accretion flow is not large. However, the small scatter in the relationship between the accretion and Br γ luminosities suggests that these considerations are not important or that they fortuitously cancel for the HAeBe stars observed by van den Ancker (2004).

Assuming the empirical relationship presented by van den Ancker (2004) is valid for our sample, we can use the accretion luminosity to infer the stellar accretion rate,

$$L_{acc} = \frac{G\dot{M}M_*}{R_*} \quad (8)$$

The stellar masses and radii are given in Table 1, and the calculated accretion rates are presented in Table 3.

If accretion heating is a dominant source of heating in the disk atmosphere, one might expect a trend between the CO emission luminosity and the accretion rate (or equivalently the flux of CO emission and Br γ). However, in addition to the caveats we mentioned before it is also important to note that the Br γ luminosities of some of our sources are indicative of accretion rates $<10^{-7}$ $M_{\odot}$ $yr^{-1}$, and rates this small were not calibrated. Also our sample spans a large range of spectral types (A7-B2). It is not clear whether the sample measured by van den Ancker (2004) spans a similar range or is more restricted. Given the various uncertainties that relate the flux of Br γ to the accretion rate, studying objects over only an order of magnitude in the accretion rate is not enough to know whether or not there is a trend. What we do find is that usually the CO P30 line is present when Br γ is observed in emission, and the CO P30 line is absent when Br γ is not observed in emission (Fig. 15). The non-detections are consistent with the cluster of data points at high accretion rate/high CO luminosity if they are linearly correlated.

There are two exceptions to the association between the accretion rate and CO line luminosity, SAO185668 and Elias 2-22b. SAO 185668 reveals a strong Br γ line but no CO emission. It is a HB3e with a large inner hole in the dust distribution (Malfait et al. 1998). Elias 2-22b does not reveal Br γ in emission, but does reveal strong CO emission. Elias 2-22 (DoAr 24E) is an IMTTS binary separated by 2" (Prato, Mathieu, & Simon 2003). The primary is modestly extinguished ($A_V \sim 6$ magnitudes), and the secondary is more heavily reddened (K-L=2.2). It is not clear how much of the K-L excess is due to extinction and how much is due to warm dust in the circumstellar disk. If all of the near infrared excess is due to reddening by interstellar dust (such that the intrinsic color is K-L=0) this implies a foreground extinction of ~40 magnitudes. Adopting $A_V = 40$ magnitudes and a normal interstellar reddening law implies $L_{Br\gamma} < 5 \times 10^{-5}$ $L_{\odot}$ and the stellar accretion rate is less than $4 \times 10^{-9}$ $M_{\odot}$ $yr^{-1}$. The upper limit is much lower than the luminosity measured from the other sources that reveal CO emission, but it is not particularly low compared to the accretion rate of classical T Tauri stars that reveal CO emission (Najita et al. 2003).

### 5.3 The inner hole around HD 141569

An interesting case among the sources in the sample is that of HD141569 where the Br γ luminosity indicates that it is accreting at nearly $10^{-8}$ $M_{\odot}$ $yr^{-1}$. Merín et al. (2004), however, place a limit of $10^{-11}$ $M_{\odot}$ $yr^{-1}$ on the accretion rate based on the low optical depth of the inner disk and the assumption that the gas-to-dust ratio is 100. Given the large inner hole inferred by mid-infrared imaging (Marsh et al. 2002), an accretion rate of $10^{-8}$ $M_{\odot}$ $yr^{-1}$ would be surprising but not unprecedented. For example, DM Tau lacks a significant NIR excess yet has a large accretion rate ( $\dot{M} = 2 \times 10^{-9} M_{\odot}\, yr^{-1}$; Calvet et al. 2005).

Several scenarios might account for a star surrounded by a transitional disk with a substantial accretion rate (Najita, Strom & Muzerolle 2006b). One scenario is that the grains have agglomerated into planetesimals such that the inner disk is gas-rich and dust-poor. A second possibility is that a giant planet of sufficient mass has formed to open a gap in the inner disk (e.g. Lin & Papaloizou 1993). Gaps in the inner disk have been inferred from models of SEDs, and are often thought to imply the presence of (proto)planetary companions (Marsh & Mahoney 1992; Calvet et al. 2002; Rice et al. 2003; D'Alessio et al. 2005; Calvet et al. 2005; Quillen et al. 2005). Material beyond the gap can cross the gap via streams that cross the orbit of the planet (Lubow et al. 1999). Estimates of the fraction of material that makes it past the planet to replenish the inner disk range from 10% (Lubow & D'Angelo 2006) to 50% (Lubow et al. 1999). A third possibility is that gas/grain interactions lead to a gas-rich/optically thin inner disk (Takeuchi & Artymowicz 2001). Each of these scenarios could result in an accreting transitional disk. However, these scenarios do not appear to apply in the case of HD 141569 because there is no substantial gas inward of 9-15 AU, as we showed in §4.

An alternative interpretation of the Br $\gamma$ emission from HD 141569 is that it is not caused by accretion and the stellar accretion rate is indeed $\leq 10^{-11}$ $M_{\odot}$ $yr^{-1}$ as suggested by Merín et al. (2004). If this is the case, the Br $\gamma$ profile may result from the same mechanism that is responsible for the Br $\gamma$ emission in shell stars; a compact disk formed by the ejection of material from a star rotating near its break-up velocity (see §3.1). If accretion of circumstellar material onto HD 141569 has ceased, then there are several possible hypotheses for the origin of its transitional SED.

One possibility is a variation of the second scenario mentioned above. In the case of a giant planet that grows to 5-10 times the gap-opening mass, accretion from the outer disk past the planet effectively ceases (Lubow et al. 1999). Thus the CO data and SED are consistent with the presence of a supra-Jovian mass planet in the inner disk. An alternative possibility is that the inner clearing is due to photo-evaporation (Clarke et al. 2001) as suggested for HD 141569 by Goto et al. (2006). In this scenario, the surface of the disk is heated to $T>10^4$ K by EUV photons (13.6 eV-100 eV photons) such that the thermal velocity of the gas is greater than the escape velocity from the star (Hollenbach et al. 2000). When the mass loss rate exceeds the disk accretion rate, the disk interior to this point is no longer replenished and an inner hole forms.

There are a few challenges to the latter interpretation, however. In the case of O and early B stars, the photoevaporation is dominated by the EUV radiation of the star[4]. For a B9.5/A0 star such as HD 141569 that part of the spectrum is relatively faint and of secondary importance for disk heating by comparison with the far ultraviolet flux of the star (FUV, 10 eV—13.6 eV). According to Adams et al. (2004), the FUV is effective at heating the disk surface to temperatures of only 100–3000 K, primarily by exciting electronic transitions of $H_2$. Those transitions are located blueward of Lyman $\alpha$, where late B/early A stars, including HD 141569, do not generate much flux (Martin-Zaidi et al. 2005; Jonkheid et al. 2006). Thus the gas at the disk surface would be expected to fall at the lower end of the temperature range suggested by Adams et al. (2004) and would restrict the rate of photoevaporation from the disk. Detailed modeling is necessary to determine whether it is feasible to achieve higher temperatures at the surface of the disk and move the photoevaporation radius into 6-15AU.

An additional challenge for the photoevaporation interpretation is the large column density of the HD141569 disk at the ostensible photoevaporation radius. The column density of $\Sigma(10AU)=3$ g $cm^{-2}$ inferred by Merín et al. (2004), based on SED fitting, is ~30 times larger than the column density at which an inner hole would form in the Clarke et al. (2001) photoevaporation model for a lower mass T Tauri star. If photoevaporation has created an inner hole in the HD141569 disk, the high column density of the disk suggests that the photoevaporation rate for that star (~$10^{-8}$ $M_{\odot}$ $yr^{-1}$) is significantly larger than the photoevaporation rate expected for T Tauri stars ($4\times10^{-10}$ $M_{\odot}$ $yr^{-1}$). Such a large mass loss rate should result in a substantial wind that is not observed in the far ultraviolet (Martin-Zaïdi et al.

---

[4] In the case of T Tauri stars, the photoevaporation is assumed to be driven by EUV flux as well (Clarke et al. 2001). The EUV flux from a T Tauri stars is thought to originate from coronal activity, but this activity is not expected for HAeBe stars.

2005). Further study is needed to determine whether photoevaporation is a viable explanation for the inner hole in the HD141569 disk.

### 5.4 Interpreting non-detections of CO emission

What constraints can we place on the gas content of disks without CO detections? In contrast to T Tauri stars, HAeBe stars have a significant UV continuum from 0.13-0.23 μm, so UV fluorescence can be an important excitation mechanism that so far has been mostly unexplored. To constrain the distribution of gas in the inner disk, we apply our modeling of v=2-1 R9 CO line (§4) to the transitional sources from which we do not detect the v=1-0 P30 CO emission (HD149914, HD 38087, SAO 185668; Fig. 16; Table 5)[5]. Each of these stars has an SED similar to the SED of HD 141569 in that they lack a NIR excess and have a MIR excess beginning at about 10 μm. Malfait et al. (1998) have modeled the SED of these stars and conclude that the inner radius of HD 38087 and SAO 185668 is 8AU and 29AU respectively. They do not report an inner radius for HD 149914; however, its MIR excess and spectral type is similar to HD 141569 so we adopt the same inner dust radius. To determine whether we can rule out a gas-rich/dust-poor region in these disks similar to what is observed around HD 141569, we apply our UV fluorescence model to these stars.

We use a synthetic UV spectrum (Kurucz 1993) appropriate for the spectral type of each star and calculate the expected flux of CO emission. Given the similarities of the disks, we make the same assumptions as in the case of HD 141569 where $T_{CO}$=200K, b=2 km $s^{-1}$, and 6≤R≤40 AU. Adopting the distances and the stellar masses listed in Table 1, we find that the flux of the *v*=2-1 R9 line should be $3.5x10^{-15}$, $5.7x10^{-15}$, and 7.3 x $10^{-16}$ erg/s/$cm^2$ from HD149914, HD 38087, SAO 185668 respectively, while the observational constraints place 3σ upper limits of $6.4x10^{-15}$, $3.0x10^{-15}$, and 4.4 x $10^{-15}$ erg/s/$cm^2$ respectively. Thus we can rule out a gas disk around HD 38087 similar to that around HD 141569. If there is a flared gaseous disk around HD 38087 it must originate beyond 20 AU from the star. The data on HD 149914 and SAO 185668 are not sensitive enough to rule out a CO disk similar to HD 141569. In the case of HD 149914, the CO would have to originate in a flared disk extending into 1AU to be detectable. The data on SAO 185668 are not sensitive enough to place any constraint on the presence of CO around this star.

## 6. Conclusions

There are two important results from this study. First, in the case of HAeBe stars, thermally excited CO closely traces dust in the inner disk. Further work is necessary to determine the degree to which Br γ is a reliable probe of the accretion rate of HAeBe stars and whether there is a connection between the presence of thermally excited CO emission and the mass accretion rate. In general, we can say that Br γ and CO emission are tightly associated. Second, UV fluoresced CO in optically thin disk systems may provide an additional method to constrain the mass and timescale of depletion of gas in the inner disk regions of slightly more evolved systems. The results provide a computational tool that will allow us to calculate the expected CO line luminosity from gas in disk holes given the UV flux and distance to the star. The quantity of gas in the dust-depleted region is important in constraining the time available for gas accretion in the planetary formation process.

---

[5] Elias 2-22a is also a source from which we do not detect CO emission, but it is a 2.3$M_{\odot}$ K0 IMMTS (Prato et al. 2003) with a weak stellar UV continuum. The UV photons needed to fluoresce any circumstellar CO would have to originate from an accretion flow, but that has not been observed. Since the UV field is unknown, it is not possible to place an upper limit on the flux of UV fluoresced CO.

We have demonstrated that UV fluorescence of CO can account for the observed ro-vibrational spectrum observed from HD 141569. The velocity profile of the gas is consistent with the spatial profile of the gas inferred by Goto et al. (2006) assuming a mass of 2.0 $M_{\odot}$. Our results indicate that there is minimal gas and dust interior to 6 AU otherwise the observed CO would be shielded from fluorescence. The relatively sharp drop off in surface density of the disk interior to 6-15 AU, combined with the large column density of the outer disk, is suggestive of the presence of a massive planet at the inner edge of the gas disk.

**Acknowledgements**

The authors wish to thank Jason Aufdenberg for running stellar models of Br γ for HD 141569 and HD 158643 and helpful discussions. The data presented herein were obtained at the W.M. Keck Observatory, which is operated as a scientific partnership among the California Institute of Technology, the University of California and the National Aeronautics and Space Administration. The Observatory was made possible by the generous financial support of the W.M. Keck Foundation. S.D.B's work was performed under contract with the Jet Propulsion Laboratory (JPL) funded by NASA through the Michelson Fellowship Program. JPL is managed for NASA by the California Institute of Technology.

Table 1. Journal of Observations

| (1) Star | (2) SpT | (3) Dist[a] pc | (4) $M_*$[a] $M_\odot$ | (5) $R_*$[a] $R_\odot$ | (6) E(K-L)[a] | (7) K | (8) $v_*\sin(i)$ km $s^{-1}$ | (9) Setting | (10) Int Time Minutes | (11) Date |
|---|---|---|---|---|---|---|---|---|---|---|
| T Tau N | G6[a] | 177±49 | 1.9 | 2.9 | 1.03 | 5.5 | 20.1±1.2[a] | MW-1 | 8 | 3/23/02 |
| | | | | | | | | K | 20 | 2/28/04 |
| HD31293 | A0[b] | 144±21 | 3.2 | 2.7 | 1.07 | 4.3 | 80±5[b] | MW-1 | 4 | 3/21/02 |
| (AB Aur) | | | | | | | | K | 4 | 3/21/02 |
| HD38087 | B5[b] | 199±75 | 5.9 | | -0.08 | 7.2 | | MW-1 | 20 | 3/20,21/02 |
| | | | | | | | | K | 12 | 3/21/02 |
| HD 250550 | B7[c] | 606±367 | 3.6 | 2.5 | 1.04 | 6.6 | 110±9[b] | MW-1 | 12 | 3/22/02 |
| | | | | | | | | K | 4 | 3/22/02 |
| HD259431 | B2[c] | 290±84 | 4.4 | 2.4 | 1.36 | 5.7 | 90±8[b] | MW-1 | 20 | 3/23/02 & 3/19/03 |
| | | | | | | | | K | 8 | 3/23/02 |
| HD58647 | B9[b] | 277±77 | 3.0 | 2.8 | 1.4 | 5.4 | | MW-1 | 14 | 3/21/02 |
| | | | | | | | | K | 16 | 3/21/02 |
| HD141569 | A0[b] | 108±6[b] | 2.0 | 1.7 | 0.21 | 6.8 | 236±9[c] | MW-1 | 16 | 3/21/02 |
| | | | | | | | | MW-2 | 40 | 3/21/02 & 3/19/03 |
| | | | | | | | | MW-3 | 20 | 3/19/03 |
| | | | | | | | | MW-4 | 20 | 3/19/03 |
| | | | | | | | | K | 16 | 3/21/02 |
| HD144668 | A7[b] | 208±31 | 1.8 | 1.9 | 1.06 | 4.3 | 180±50[b] | MW-1 | 4 | 3/22/02 |
| (HR 5999) | | | | | | | | K | 4 | 3/22/02 |
| HD149914 | B9.5[b] | 165±27 | 3.0 | 2.8 | 0.08 | 5.7 | | MW-1 | 8 | 3/23/02 |
| | | | | | | | | K | 8 | 3/21/02 |
| HD158643 | B9.5[b] | 131±15 | 2.9 | 2.7 | 0.90 | 4.3 | 267±5[c] | MW-1 | 4 | 3/22/02 |
| (51Oph) | | | | | | | | K | 12 | 3/21/02 |
| SAO 185668 | B3[b] | 700[d] | 7.6 | 4.2 | 0.56 | 8.5 | | MW-1 | 8 | 3/23/02 |
| | | | | | | | | K | 8 | 3/21/02 |
| HD163296 | A1[b] | 122±15 | 2.3 | 2.1 | 1.18 | 4.7 | 120±30[d] | MW-1 | 8 | 3/23/02 |
| | | | | | | | | K | 8 | 3/23/02 |
| Elias2-22a | K0[d] | 140[e] | 2.3 | 3 | 0.4 | 7.1 | | MW-1 | 12 | 3/23/02 |
| (DoAr 24E) | | | | | | | | K | 8 | 3/21/02 |
| Elias2-22b | … | 140[e] | 1 | 1.5 | <2.2 | 8.1 | | MW-1 | 12 | 3/23/02 |
| (DoAr 24Eb) | | | | | | | | K | 8 | 3/21/02 |

(2) a-Calvet et al. 2004; b-Malfait et al. 1998; c-Hillenbrand et al. 1992; d-Prato et al. 2003

(3) a-Distances determined from Hipparcos measurements presented by SIMBAD except as noted; b - Merín et al. 2004; d - SAO 185668: from distance modulus given $m_V$=9.1 and $A_V$=1.8 (Malfait et al. 1998); e - Elias 2-22: from distance modulus given $m_J$=12.1 and $A_J$=1.8 (Gras-Velázquez & Ray 2005)

(4,5) a-The mass and radius of a star were inferred from the spectral type assuming that the it was on the main-sequence, except for T Tau N (Calvet et al. 2004), HD 141569 (Merín et al. 2004) and Elias 2-22a (Prato et al. 2003). Since nothing is known about the nature of Elias 2-22b, we have adopted 1$M_\odot$.

(6,7) a-The infrared magnitudes were taken from Malfait et al. (1998) with the exceptions: T Tau (Beck et al. 2004); HD 250550 and HD 259431 (Hillenbrand et al. 1992); Elias 2-22 (Prato et al. 2003)

(8) a - Hartmann et al. 1986; b – Böhm & Catala 1995; c – Dunkin et al. 1997b; d – Finkenzeller 1985

Table 2. Instrument Settings

| Setting | Order | Wavenumber range |
|---|---|---|
| M-wide-1(61.12, 37.05)[1] | 14 | 1885-1855 |
| | 15 | 2019-1987 |
| | 16 | 2153-2118 |
| M-wide-2 (62.4, 37.25) | 14 | 1857-1829 |
| | 15 | 1889-1960 |
| | 16 | 2120-2090 |
| M-wide-3 (64.28, 36.70) | 15 | 1962-1935 |
| | 16 | 2091-2063 |
| | 17 | 2221-2191 |
| M-wide-4 (65.88,36.86) | 15 | 1935-1910 |
| | 16 | 2063-2037 |
| | 17 | 2191-2166 |
| K1(62.53/35.55) | 32 | 4252-4198 |
| | 33 | 4384-4318 |
| | 34 | 4516-4448 |
| | 35 | 4648-4578 |
| | 36 | 4781-4708 |
| | 37 | 4912-4837 |
| | 38 | 5010-4967 |

[1]The numbers refer to the cross disperser and grating setting.

Table 3. Measured Parameters

| *(1)* | *(2)* | *(3)* | *(4)* | *(5)* | *(6)* | *(7)* | *(8)* | *(9)* | *(10)* |
|---|---|---|---|---|---|---|---|---|---|
| Star | $V_{rad}$ | $V_{rad}$ Br γ | $V_{rad}$ CO | FWZI Br γ | $L_{Br\gamma}$ | $\dot{M}$ | FWHM v=1-0 P30 CO | HWZI v=1-0 P30 CO | Luminosity v=1-0 P30 CO |
| | km $s^{-1}$ | km $s^{-1}$ | km $s^{-1}$ | km $s^{-1}$ | $10^{-4}$ $L_{\odot}$ | $10^{-7}$ $M_{\odot}$ $yr^{-1}$ | km $s^{-1}$ | km $s^{-1}$ | $10^{22}$ W |
| T Tau N | 19.1[a] | 19±4 | 17±4 | 300 | 3.2 | .51[a] | 40 | 75 | 7.1 |
| HD31293 (AB Aur) | 9[b] | 9±4 | 11±2 | 350 | 18.9 | 2.6 | 15 | 40 | 1.8 |
| HD38087 | 33[c] | 30±4 | -- | -- | 0.42 $cm^{-1}$[a] | -- | -- | -- | <0.05 |
| HD 250550 | 16[d] | 23±4 | 17±3 | 500 | 66.2 | 10.5 | 30 | 40 | 20.2 |
| HD259431 | 17[d] | 19±4 | 23±8 | 350 | 38.1 | 4.1 | 60 | 77 | 3.7 |
| HD58647 | -- | 2±4 | 16±8 | 400 | 21.8 | 3.5 | 40 | 50 | 1.3 |
| HD141569 | -6[d] | -22±8 | -7±2 | 600 | 1.1 | 0.08 | 12 | -- | 0.005[a] |
| HR 5999 | 26[b] | -5±4 | 35±8 | 475 | 45.8 | 10.0 | 45 | 47 | 2.0 |
| HD149914 | -- | -18±4 | -- | -- | 1.34 $cm^{-1}$[a] | -- | -- | -- | <0.07 |
| HD158643 (51 Oph) | -20[d] | -20±8 | -20±8 | 550 | 10.4 | 1.4 | 250 | 200 | 4.5 |
| SAO 185668 | -- | 61±4 | -- | 350 | 2.14 | 0.11 | -- | -- | <0.08 |
| HD163296 | -4[b] | -27±4 | -19±4 | 800 | 7.43 | 0.88 | 40 | 50 | 0.75 |
| Elias2-22a (DoAr 24Ea) | -- | -- | -- | -- | <0.11 | <0.006 | | | <0.03 |
| Elias2-22b (DoAr 24Eb) | -- | -- | -17±4 | -- | <$9.6\times10^{-3}$ [b] | <$3\times10^{-4}$ | 50 | 82 | 1.5[b] |

(2) a - Hartmann et al. 1986; b-Acke et al. 2005; c-Wilson 1953 (via SIMBAD); d-Finkenzeller & Jankovics 1984; e-Dunkin et al. 1997b

(6) a – Br γ is in absorption, thus the equivalent width is given in wavenumbers ($cm^{-1}$). b- This source has not been corrected for reddening

(7) a - Calvet et al. (2004) measure the accretion rate of T Tau N from the near UV excess and find that it varies between 3-5 x $10^{-8}$ $M_{\odot}$ $yr^{-1}$.

(10) a - Extrapolated from lower lying lines; b-This luminosity has not been corrected for reddening.

Table 4
Measurements of CO Emission from HD 141569
Rest Position, Observed Position, and Integrated Flux of Ro-Vibrational CO Transitions

| Isotopomer | v" | J" | v' | J' | $\tilde{\nu}_{res}$ cm$^{-1}$ | $\tilde{\nu}_{obs}$ cm$^{-1}$ | $v_{rad}$ km s$^{-1}$ | F $10^{-14}$ erg s$^{-1}$ cm$^{-2}$ |
|---|---|---|---|---|---|---|---|---|
| $^{12}$CO | 0 | 11 | 1 | 12 | 2186.64 | 2186.80 | -22.1 | 0.63±0.34 |
| $^{12}$CO | 0 | 10 | 1 | 11 | 2183.22 | 2183.45 | -31.1 | 0.64±0.34 |
| $^{12}$CO | 0 | 9 | 1 | 10 | 2179.77 | 2180.03 | -35.5 | 0.83±0.34 |
| $^{12}$CO | 0 | 8 | 1 | 9 | 2176.28 | 2176.50 | -29.8 | 0.85±0.17 |
| $^{12}$CO | 0 | 7 | 1 | 8 | 2172.76 | 2173.01 | -34.7 | 1.05±0.17 |
| $^{12}$CO | 0 | 6 | 1 | 7 | 2169.20 | 2169.43 | -32.1 | 1.12±0.17 |
| $^{12}$CO | 0 | 5 | 1 | 6 | 2165.60 | 2165.84 | -33.1 | 1.17±0.17 |
| $^{12}$CO | 0 | 2 | 1 | 3 | 2154.60 | 2154.81 | 30.2 | 1.55±0.05 |
| $^{12}$CO | 0 | 1 | 1 | 2 | 2150.86 | 2151.05 | 26.9 | 1.34±0.07 |
| $^{12}$CO | 0 | 0 | 1 | 1 | 2147.08 | 2147.29 | 29.5 | 0.66±0.08 |
| $^{12}$CO | 0 | 1 | 1 | 0 | 2139.43 | 2139.67 | 33.8 | 0.70±0.08 |
| $^{12}$CO | 0 | 2 | 1 | 1 | 2135.55 | 2135.78 | 32.8 | 1.32±0.05 |
| $^{12}$CO | 0 | 5 | 1 | 4 | 2123.70 | 2123.93 | 32.7 | 1.85±0.06 |
| $^{12}$CO | 0 | 6 | 1 | 5 | 2119.68 | 2119.91 | 32.0 | 2.02±0.22 |
| $^{12}$CO | 0 | 7 | 1 | 6 | 2115.63 | 2115.87 | 34.0 | 2.11±0.24 |
| $^{12}$CO | 0 | 8 | 1 | 7 | 2111.54 | 2111.76 | 31.1 | 2.05±0.32 |
| $^{12}$CO | 0 | 9 | 1 | 8 | 2107.42 | 2107.65 | 32.8 | 1.16±0.10 |
| $^{12}$CO | 0 | 14 | 1 | 13 | 2086.32 | 2086.55 | 33.1 | 0.44±0.08 |
| $^{12}$CO | 0 | 15 | 1 | 14 | 2082.00 | 2082.21 | 30.0 | 0.38±0.12 |
| $^{12}$CO | 0 | 17 | 1 | 16 | 2073.26 | 2073.45 | 26.6 | 0.24±0.09 |
| $^{12}$CO | 0 | 20 | 1 | 19 | 2059.91 | 2060.11 | 28.4 | 0.05±0.05 |
| $^{12}$CO | 1 | 10 | 2 | 11 | 2156.36 | 2156.60 | 34.0 | 0.71±0.03 |
| $^{12}$CO | 1 | 9 | 2 | 10 | 2152.94 | 2153.17 | 31.8 | 0.73±0.03 |
| $^{12}$CO | 1 | 8 | 2 | 9 | 2149.49 | 2149.70 | 29.5 | 0.87±0.03 |
| $^{12}$CO | 1 | 7 | 2 | 8 | 2146.00 | 2146.22 | 30.4 | 1.01±0.03 |
| $^{12}$CO | 1 | 6 | 2 | 7 | 2142.47 | 2142.70 | 31.3 | 0.97±0.03 |
| $^{12}$CO | 1 | 5 | 2 | 6 | 2138.91 | 2139.14 | 32.7 | 1.19±0.03 |
| $^{12}$CO | 1 | 2 | 2 | 3 | 2128.01 | 2128.23 | 31.4 | 1.08±0.11 |
| $^{12}$CO | 1 | 1 | 2 | 2 | 2124.31 | 2124.52 | 30.8 | 1.56±0.38 |
| $^{12}$CO | 1 | 0 | 2 | 1 | 2120.57 | 2120.81 | 35.1 | 0.65±0.15 |
| $^{12}$CO | 1 | 1 | 2 | 0 | 2112.98 | 2113.21 | 33.0 | 0.66±0.18 |
| $^{12}$CO | 1 | 2 | 2 | 1 | 2109.14 | 2109.37 | 32.9 | 0.90±0.06 |
| $^{12}$CO | 1 | 5 | 2 | 4 | 2097.39 | 2097.62 | 32.2 | 1.00±0.07 |
| $^{12}$CO | 1 | 7 | 2 | 6 | 2089.39 | 2089.64 | 35.7 | 1.20±0.23 |
| $^{12}$CO | 1 | 8 | 2 | 7 | 2085.34 | 2085.57 | 32.7 | 1.14±0.18 |
| $^{12}CO$ | 1 | 9 | 2 | 8 | 2081.26 | 2081.45 | 27.1 | 1.33±0.21 |
| $^{12}CO$ | 1 | 18 | 2 | 17 | 2043.00 | 2043.20 | 29.8 | 0.25±0.08 |
| $^{12}CO$ | 2 | 12 | 3 | 13 | 2136.21 | 2136.45 | 33.0 | 0.26±0.03 |
| $^{12}CO$ | 2 | 11 | 3 | 12 | 2132.91 | 2133.13 | 31.9 | 0.37±0.03 |
| $^{12}CO$ | 2 | 10 | 3 | 11 | 2129.56 | 2129.82 | 36.7 | 0.39±0.04 |
| $^{12}CO$ | 2 | 9 | 3 | 10 | 2126.18 | 2126.41 | 33.1 | 0.48±0.03 |
| $^{12}CO$ | 2 | 7 | 3 | 8 | 2119.30 | 2119.60 | 42.3 | 0.67±0.07 |
| $^{12}CO$ | 2 | 6 | 3 | 7 | 2115.81 | 2116.03 | 30.5 | 0.74±0.06 |
| $^{12}CO$ | 2 | 5 | 3 | 6 | 2112.29 | 2112.52 | 33.0 | 0.65±0.05 |
| $^{12}CO$ | 2 | 4 | 3 | 5 | 2108.72 | 2108.96 | 33.1 | 1.00±0.22 |
| $^{12}CO$ | 2 | 1 | 3 | 2 | 2097.82 | 2098.04 | 31.7 | 0.56±0.05 |
| $^{12}CO$ | 2 | 1 | 3 | 0 | 2086.60 | 2086.81 | 29.7 | 0.46±0.08 |
| $^{12}CO$ | 2 | 5 | 3 | 4 | 2071.15 | 2071.37 | 31.3 | 0.70±0.08 |
| $^{12}CO$ | 2 | 7 | 3 | 6 | 2063.22 | 2063.45 | 33.2 | 0.69±0.08 |
| $^{12}CO$ | 2 | 8 | 3 | 7 | 2059.21 | 2059.45 | 34.5 | 1.24±0.10 |
| $^{12}CO$ | 2 | 9 | 3 | 8 | 2055.16 | 2055.37 | 31.3 | 0.39±0.12 |
| $^{12}CO$ | 2 | 10 | 3 | 9 | 2051.08 | 2051.29 | 31.5 | 0.54±0.09 |
| $^{12}CO$ | 2 | 11 | 3 | 10 | 2046.96 | 2047.19 | 34.0 | 0.57±0.09 |
| $^{12}CO$ | 2 | 13 | 3 | 12 | 2038.62 | 2038.86 | 34.8 | 0.45±0.09 |
| $^{12}CO$ | 3 | 8 | 4 | 9 | 2096.10 | 2096.31 | 30.8 | 0.35±0.05 |
| $^{12}CO$ | 3 | 5 | 4 | 6 | 2085.73 | 2085.95 | 31.8 | 0.35±0.08 |
| $^{12}CO$ | 3 | 2 | 4 | 3 | 2075.04 | 2075.28 | 35.1 | 1.18±0.12 |
| $^{12}CO$ | 3 | 1 | 4 | 2 | 2071.40 | 2071.63 | 33.2 | 0.29±0.08 |
| $^{12}CO$ | 3 | 3 | 4 | 2 | 2052.71 | 2052.92 | 32.0 | 0.77±0.10 |
| $^{12}CO$ | 3 | 7 | 4 | 6 | 2037.12 | 2037.34 | 32.0 | 1.05±0.16 |
| $^{12}CO$ | 3 | 13 | 4 | 12 | 2012.73 | 2012.97 | 35.9 | 0.26±0.03 |
| $^{12}CO$ | 3 | 14 | 4 | 13 | 2008.55 | 2008.70 | 21.9 | 0.16±0.03 |
| $^{12}CO$ | 3 | 16 | 4 | 15 | 2000.09 | 2000.32 | 35.0 | 0.14±0.03 |
| $^{12}CO$ | 3 | 17 | 4 | 16 | 1995.81 | 1996.03 | 33.7 | 0.08±0.03 |
| $^{12}CO$ | 4 | 5 | 5 | 6 | 2059.24 | 2059.45 | 29.6 | 1.26±0.10 |
| $^{12}CO$ | 4 | 4 | 5 | 5 | 2055.75 | 2055.97 | 32.3 | 0.31±0.09 |
| $^{12}CO$ | 4 | 3 | 5 | 4 | 2052.22 | 2052.45 | 33.1 | 0.80±0.10 |
| $^{12}CO$ | 4 | 1 | 5 | 2 | 2045.06 | 2045.23 | 26.0 | 0.65±0.09 |
| $^{12}CO$ | 4 | 7 | 5 | 6 | 2011.09 | 2011.30 | 31.8 | 0.31±0.03 |
| $^{12}CO$ | 4 | 8 | 5 | 7 | 2007.14 | 2007.38 | 34.8 | 0.35±0.03 |
| $^{12}CO$ | 4 | 9 | 5 | 8 | 2003.17 | 2003.36 | 28.6 | 0.33±0.03 |
| $^{12}CO$ | 4 | 11 | 5 | 10 | 1995.10 | 1995.29 | 28.3 | 0.18±0.03 |
| $^{12}CO$ | 5 | 2 | 6 | 3 | 2022.35 | 2022.56 | 32.1 | 0.25±0.03 |
| $^{12}CO$ | 5 | 2 | 6 | 1 | 2004.17 | 2004.41 | 36 | 0.45±0.03 |
| $^{12}CO$ | 5 | 3 | 6 | 2 | 2000.43 | 2000.66 | 34.6 | 0.30±0.03 |
| $^{12}CO$ | 5 | 4 | 6 | 3 | 1996.66 | 1996.87 | 31.8 | 0.30±0.03 |
| $^{13}CO$ | 0 | 12 | 1 | 13 | 2140.83 | 2141.08 | 35.1 | 0.21±0.03 |
| $^{13}CO$ | 0 | 11 | 1 | 12 | 2137.59 | 2137.80 | 30.1 | 0.14±0.03 |
| $^{13}CO$ | 0 | 9 | 1 | 10 | 2131.00 | 2131.23 | 32.4 | 0.41±0.03 |
| $^{13}CO$ | 0 | 6 | 1 | 7 | 2120.87 | 2121.09 | 30 | 0.25±0.06 |
| $^{13}CO$ | 0 | 5 | 1 | 6 | 2117.43 | 2117.64 | 29.6 | 0.45±0.06 |
| $^{13}CO$ | 0 | 4 | 1 | 5 | 2113.95 | 2114.18 | 32.7 | 0.32±0.05 |
| $^{13}CO$ | 0 | 3 | 1 | 4 | 2110.44 | 2110.68 | 33.5 | 0.36±0.05 |
| $^{13}CO$ | 0 | 2 | 1 | 3 | 2106.90 | 2107.11 | 30.8 | 0.40±0.07 |
| $^{13}CO$ | 0 | 0 | 1 | 1 | 2099.71 | 2099.92 | 29.8 | 0.26±0.05 |
| $^{13}CO$ | 0 | 2 | 1 | 1 | 2088.68 | 2088.92 | 34 | 0.32±0.09 |
| $^{13}CO$ | 0 | 3 | 1 | 2 | 2084.94 | 2085.18 | 34.6 | 0.19±0.08 |
| $^{13}CO$ | 0 | 7 | 1 | 6 | 2069.66 | 2069.88 | 32 | 0.66±0.08 |
| $^{13}CO$ | 0 | 12 | 1 | 11 | 2049.83 | 2050.05 | 31.7 | 0.30±0.09 |
| $^{13}CO$ | 1 | 13 | 2 | 14 | 2118.26 | 2118.44 | 25.6 | 0.30±0.08 |
| $^{13}CO$ | 1 | 12 | 2 | 13 | 2115.09 | 2115.29 | 28.4 | 0.25±0.10 |
| $^{13}CO$ | 1 | 8 | 2 | 9 | 2102.05 | 2102.27 | 31 | 0.25±0.05 |
| $^{13}CO$ | 1 | 7 | 2 | 8 | 2098.71 | 2098.93 | 31.7 | 0.27±0.06 |
| $^{13}CO$ | 1 | 6 | 2 | 7 | 2095.33 | 2095.56 | 32.4 | 0.76±0.39 |
| $^{13}CO$ | 1 | 3 | 2 | 4 | 2085.00 | 2085.18 | 26.7 | 0.82±0.08 |
| $^{13}CO$ | 1 | 4 | 2 | 3 | 2055.98 | 2056.22 | 34.5 | 0.58±0.09 |
| $^{13}CO$ | 1 | 6 | 2 | 5 | 2048.41 | 2048.65 | 35.5 | 0.42±0.10 |
| $^{13}CO$ | 1 | 7 | 2 | 6 | 2044.57 | 2044.80 | 33.4 | 0.25±0.10 |

Table 5. Flux of v=2-1 R9 CO emission line

| Star | SpT | E(K-L) | F (v=2-1 R9 CO) | L (v=2-1 R9 CO) |
|---|---|---|---|---|
| | | | $10^{-15}$ erg $s^{-1}$ $cm^{2}$ | $10^{24}$ erg $s^{-1}$ |
| HD38087 | B5 | -0.08 | <3.0 | <1.4 |
| HD141569 | B9.5/A0 | 0.21 | 7.3 | 0.91 |
| HD149914 | B9 | 0.08 | <6.4 | <2.1 |
| SAO 185668 | B3 | 0.56 | <4.4 | <26 |
| Elias2-22a | K0 | 0.41 | <17 | <5.2 |

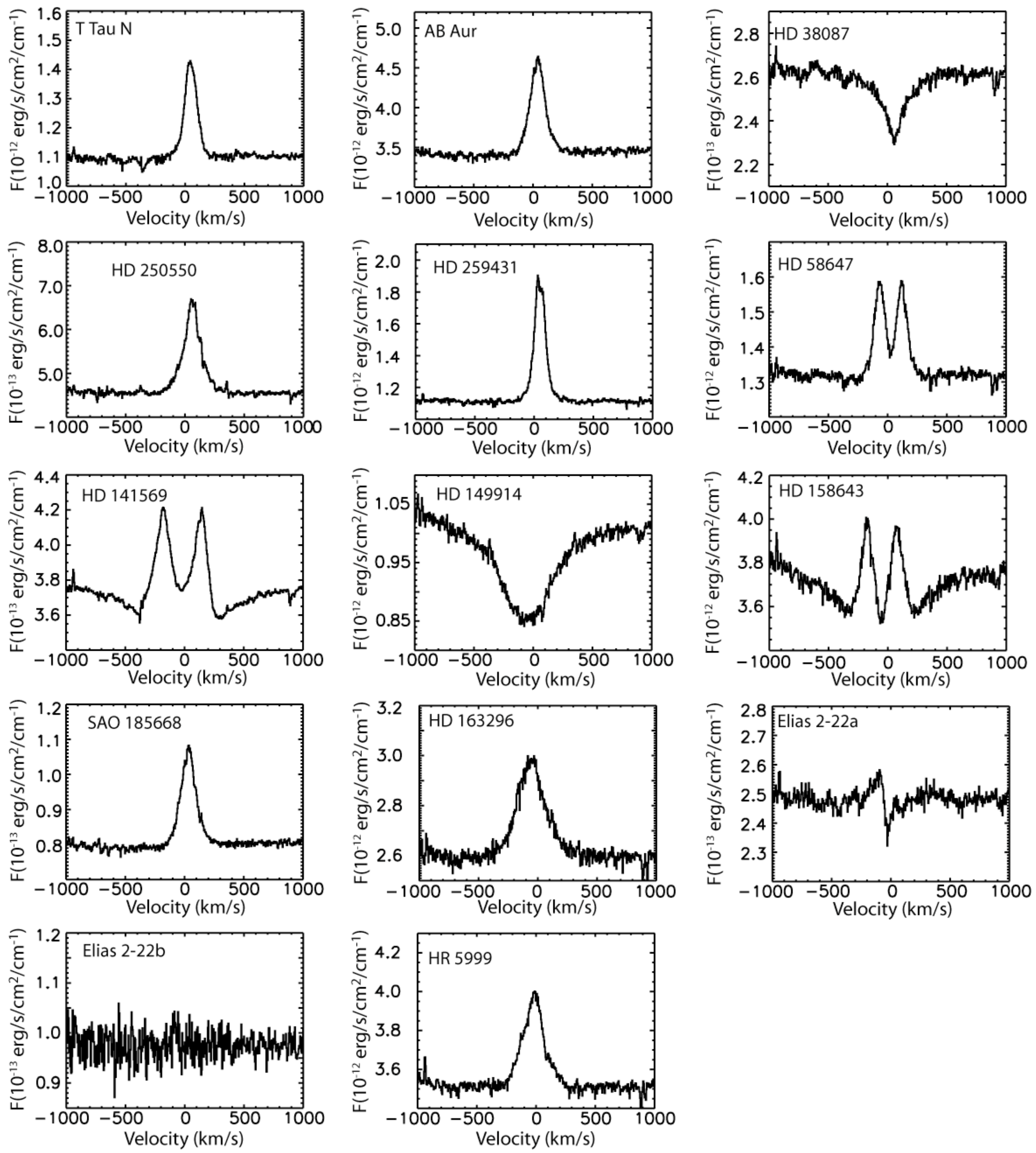

**Figure 1.** Br γ emission. We detect the Br γ transition ($\tilde{\upsilon} = 4616.55\ cm^{-1}$) from 12 of the 14 sources. The flux scale was calculated from the K-band magnitude presented in Table 1.

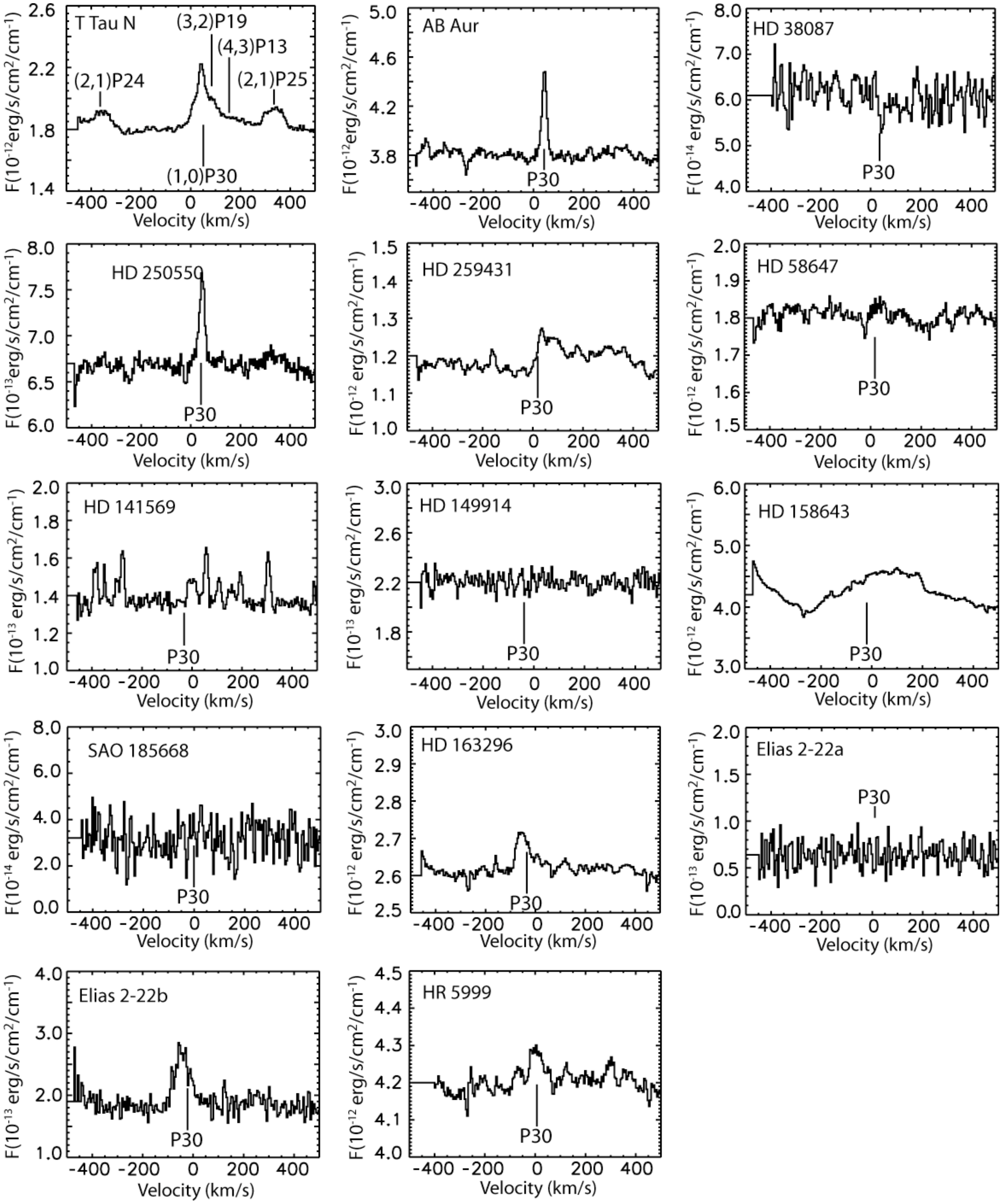

**Figure 2.** Spectra of the CO v=1-0 P30 line. We chose to focus on the P30 ($\tilde{\upsilon} = 2013.35\ cm^{-1}$) line because it is consistently unaffected by telluric absorption lines and is minimally blended with other vibrational bands of CO. We detect the P30 line in 9 out of 14 sources. The emission lines apparent in the spectrum of HD 141569 are not due to the v=1-0 vibrational branch. Emission lines from this vibrational branch fade out at P20 (see figure 6 for the complete CO spectrum of HD 141569).

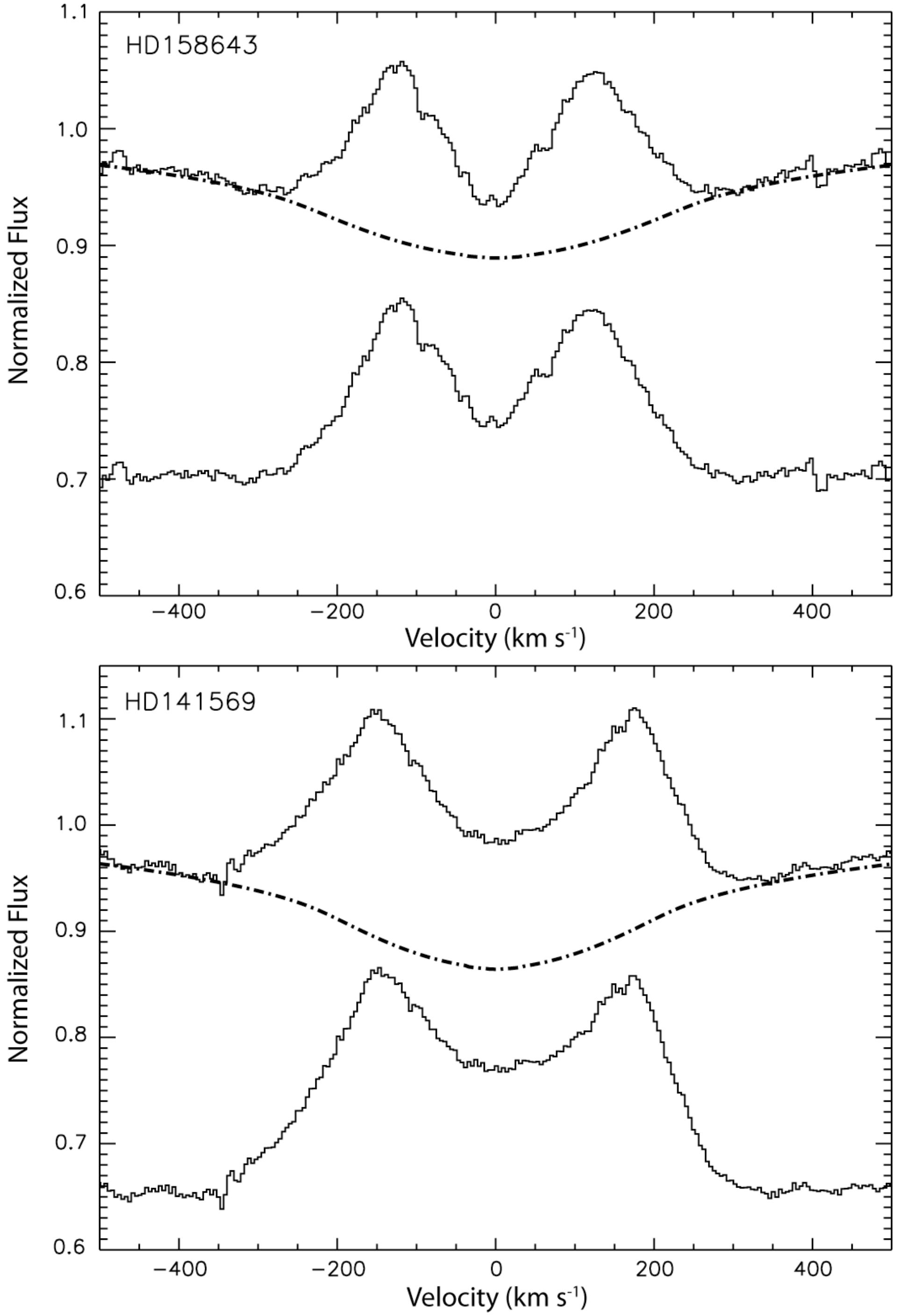

**Figure 3**. Fit of a continuum veiled theoretical profile of Br γ to the wings of the Br γ lines in HD 158643 and HD 141569 (courtesy of J. Aufdenberg). The dot-dashed line is the modeled profile of the rotationally broadened Br γ line plotted over the normalized photospheric spectrum of the star. The difference of the photospheric spectrum and the model is plotted as well and shifted up 0.7 and 0.65 units respectively.

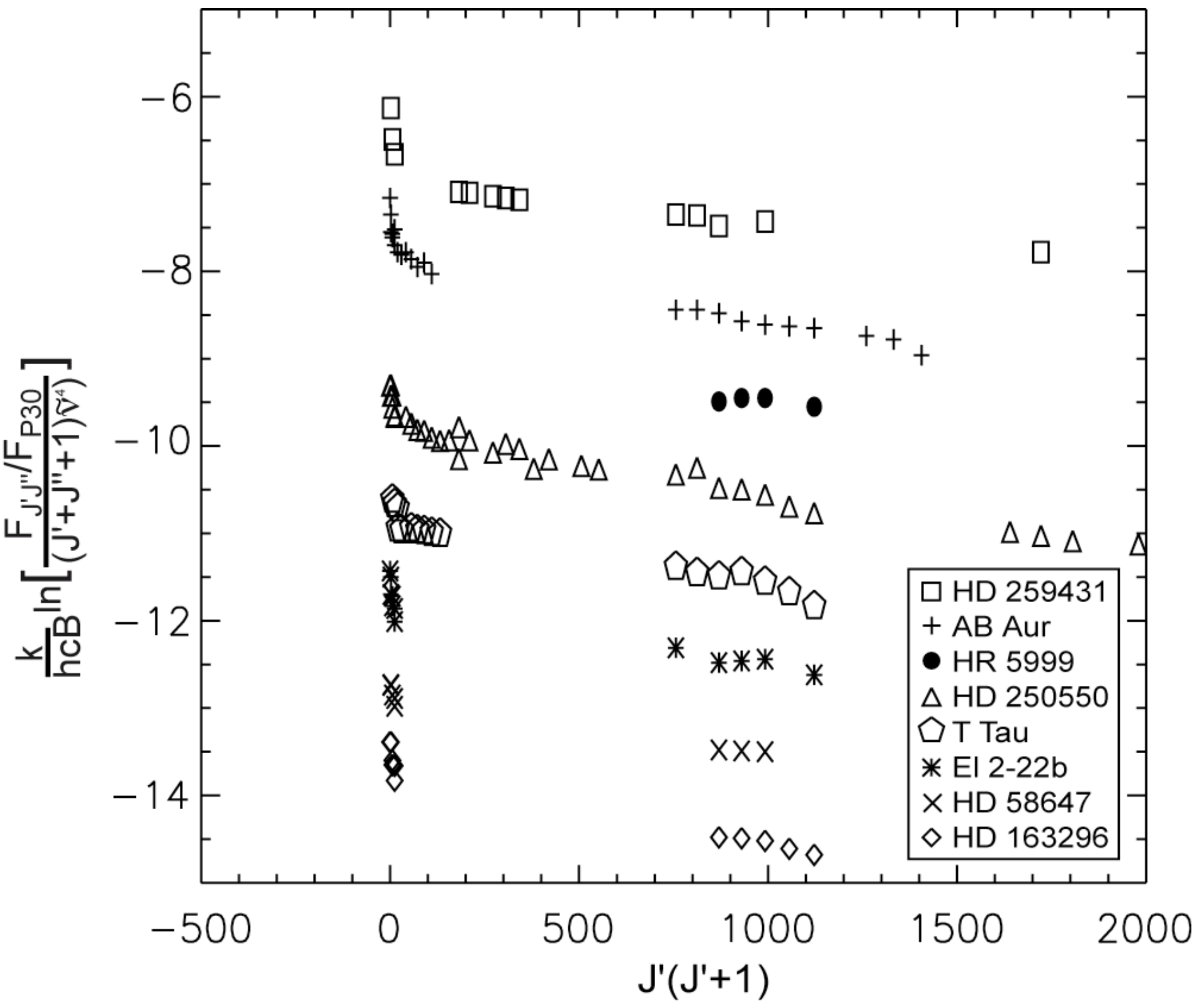

**Figure 4.** CO Excitation plot of HAeBe stars with optically thick inner disks. The $^{12}CO$ emission lines have been normalized to the flux of the v=1-0 P30 line. The population of the high-J lines indicates that the gas is hot (T>1000K). The relatively steep slope for the low-J lines and shallow slope for the high-J lines is also seen for the HAeBe stars observed by Blake & Boogert (2004). The non-linearity of the plot suggests that the gas is optically thick and/or that the beam includes gas emitting at different temperatures.

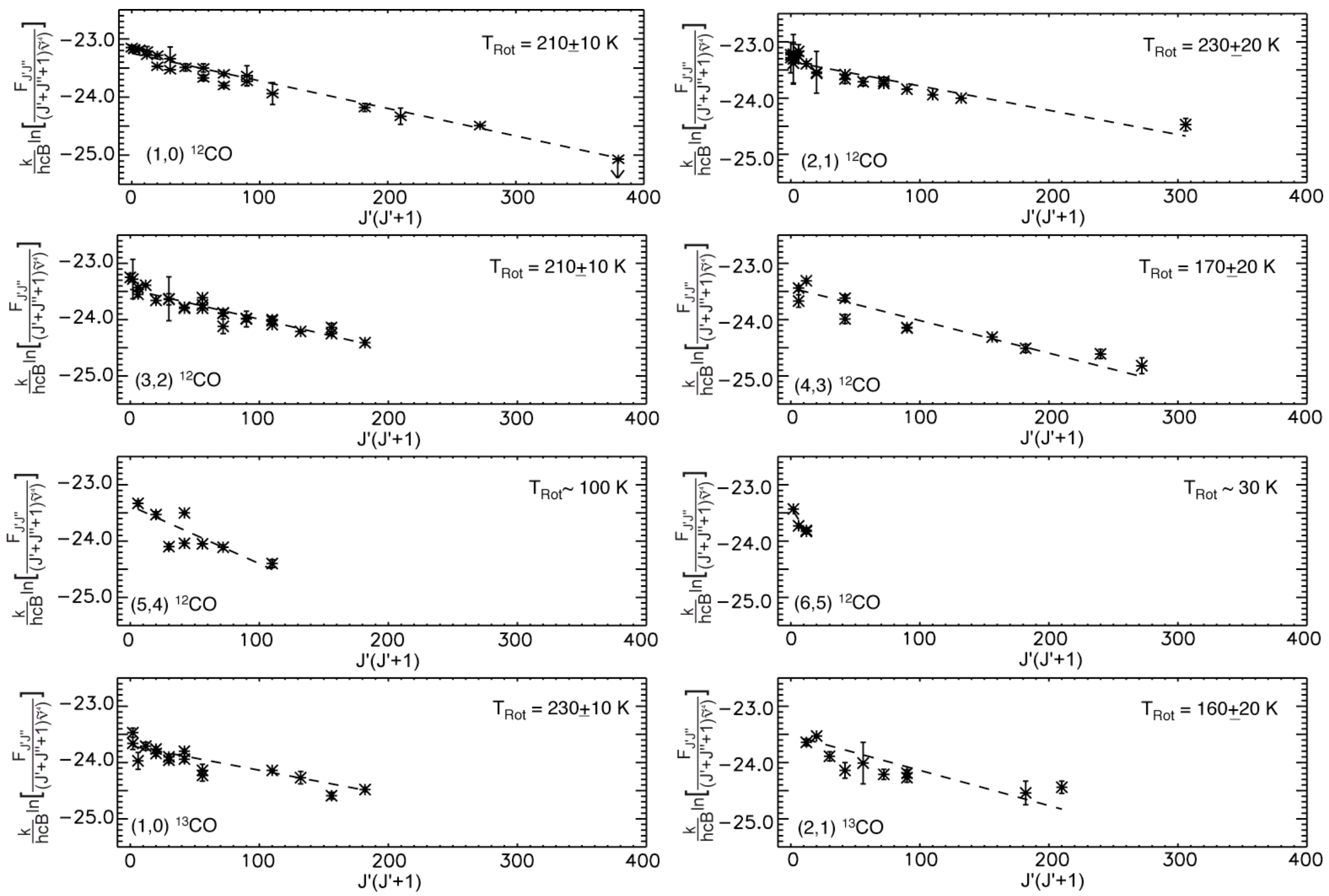

**Figure 5.** Rotational temperatures of observed vibrational levels of CO are presented for HD 141569. The gas is only 200 K, and the linearity of the points indicates that the gas is optically thin to escaping IR photons.

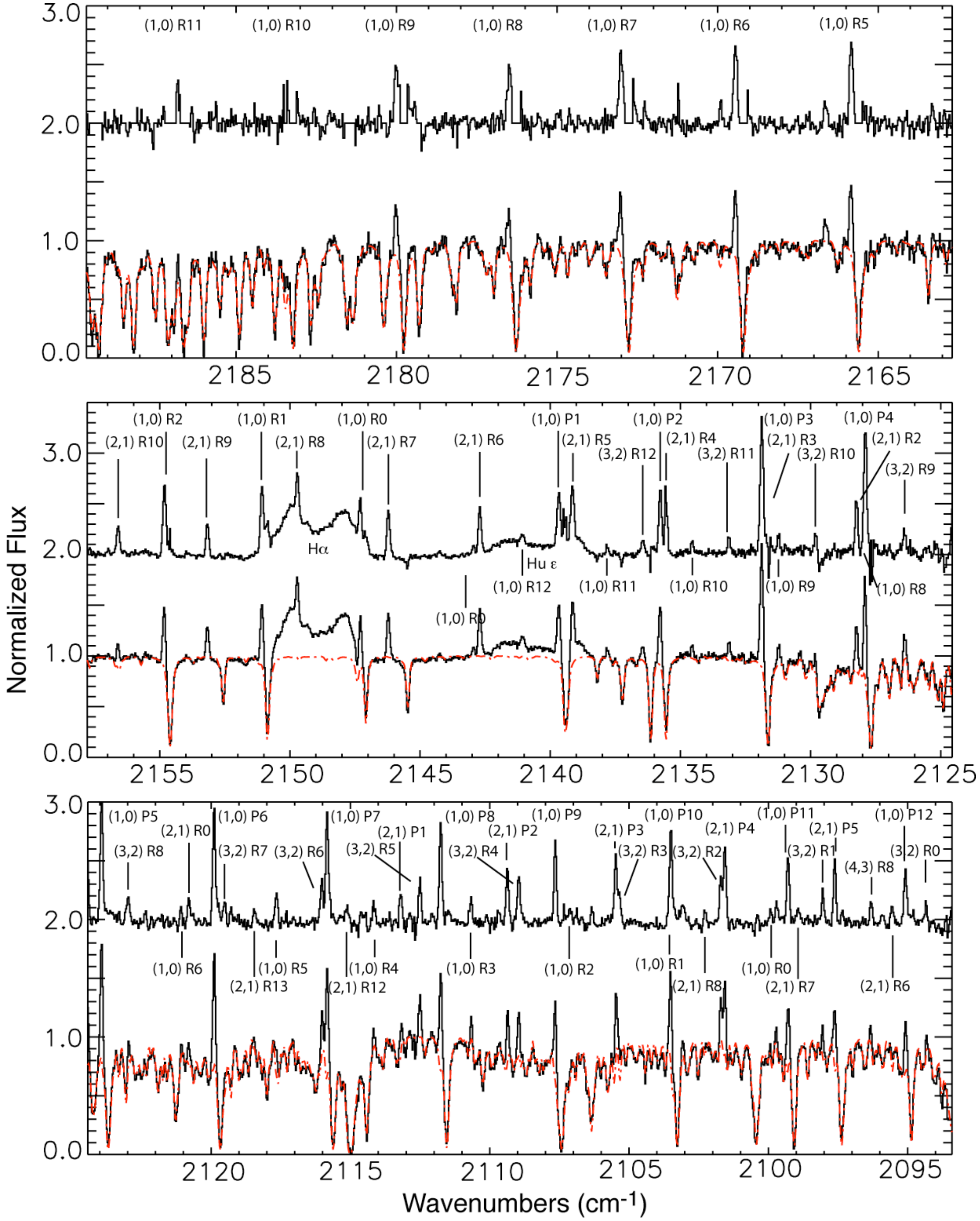

**Figure 6.** CO emission spectrum from HD 141569. $^{12}$CO emission lines from v'=1, 2, 3, 4, 5, and 6 and $^{13}$CO emission lines from v'=1 and 2 are seen. The dashed line is the atmospheric transmittance model plotted over the observed spectrum. The ratioed spectrum is plotted above.

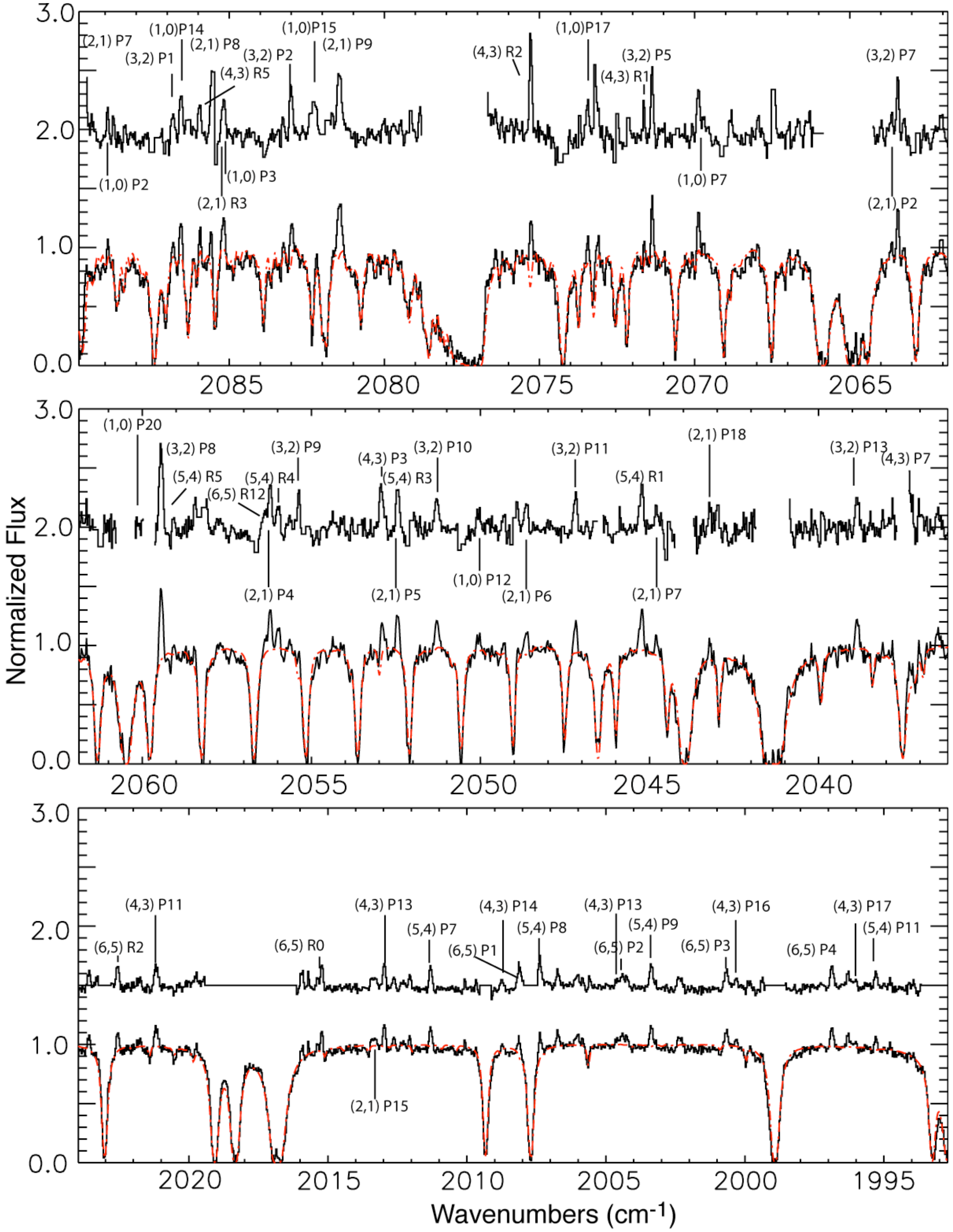

**Figure 6 continued:** Gaps in the ratioed spectrum correspond to telluric features with a transmittance less than 50%. Labels for $^{12}C^{16}O$ transitions are placed above the ratioed spectrum. Labels for $^{13}C^{16}O$ transitions are placed below the ratioed spectrum.

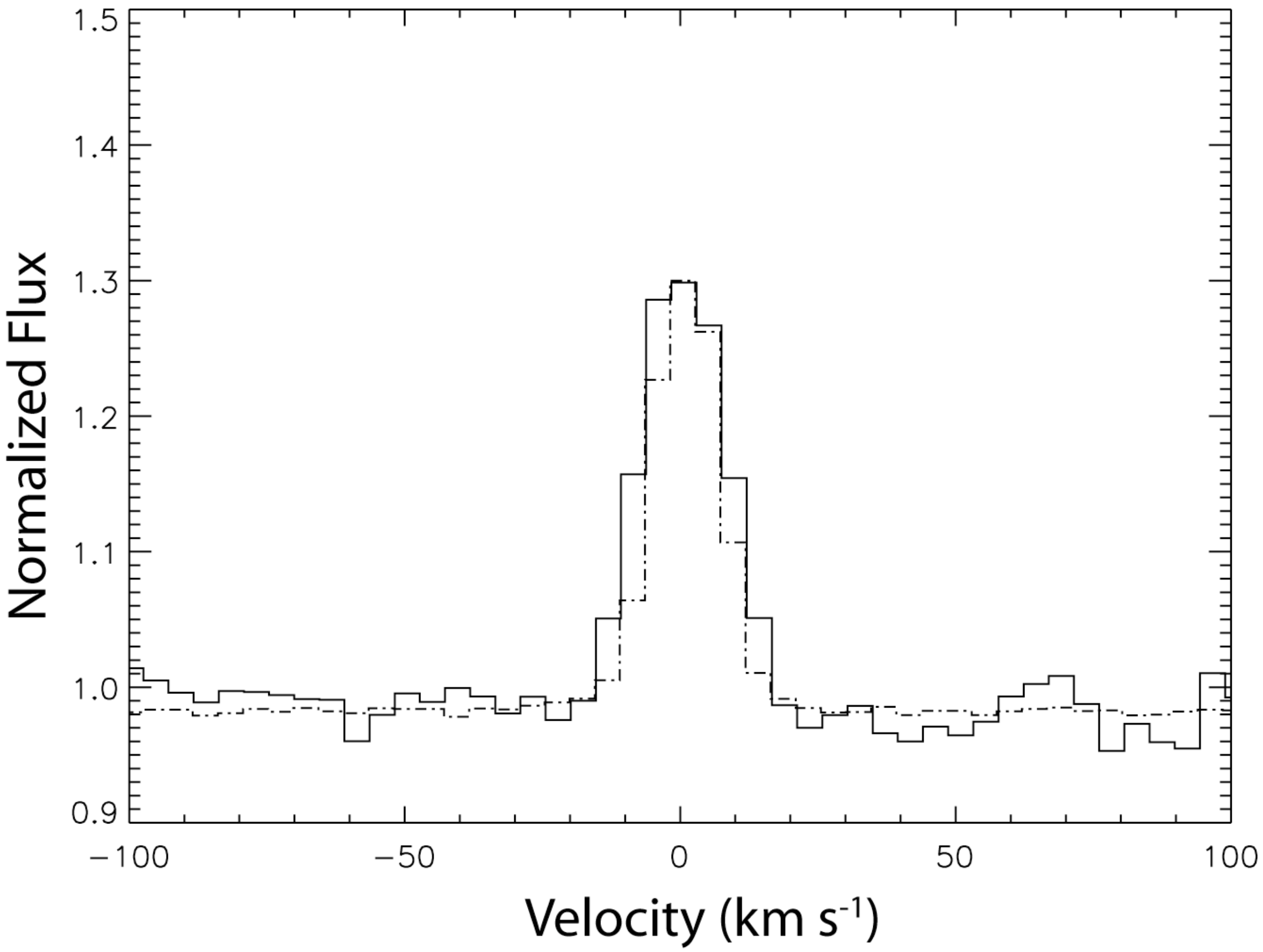

**Figure 7.** Comparison of the v=2-1 R9 line to an arc-lamp line. An unsaturated arc-lamp line at 2 μm was selected to compare with the profile of the emission line observed from HD 141569. The HWZI of the arc-lamp line is 22 km $s^{-1}$ while the HWZI of the CO emission line is ≤ 26 km $s^{-1}$. Since the inclination of the disk is 51° (Weinberger et al. 2000), the upper limit on the maximum de-projected velocity of the gas is 18 km $s^{-1}$.

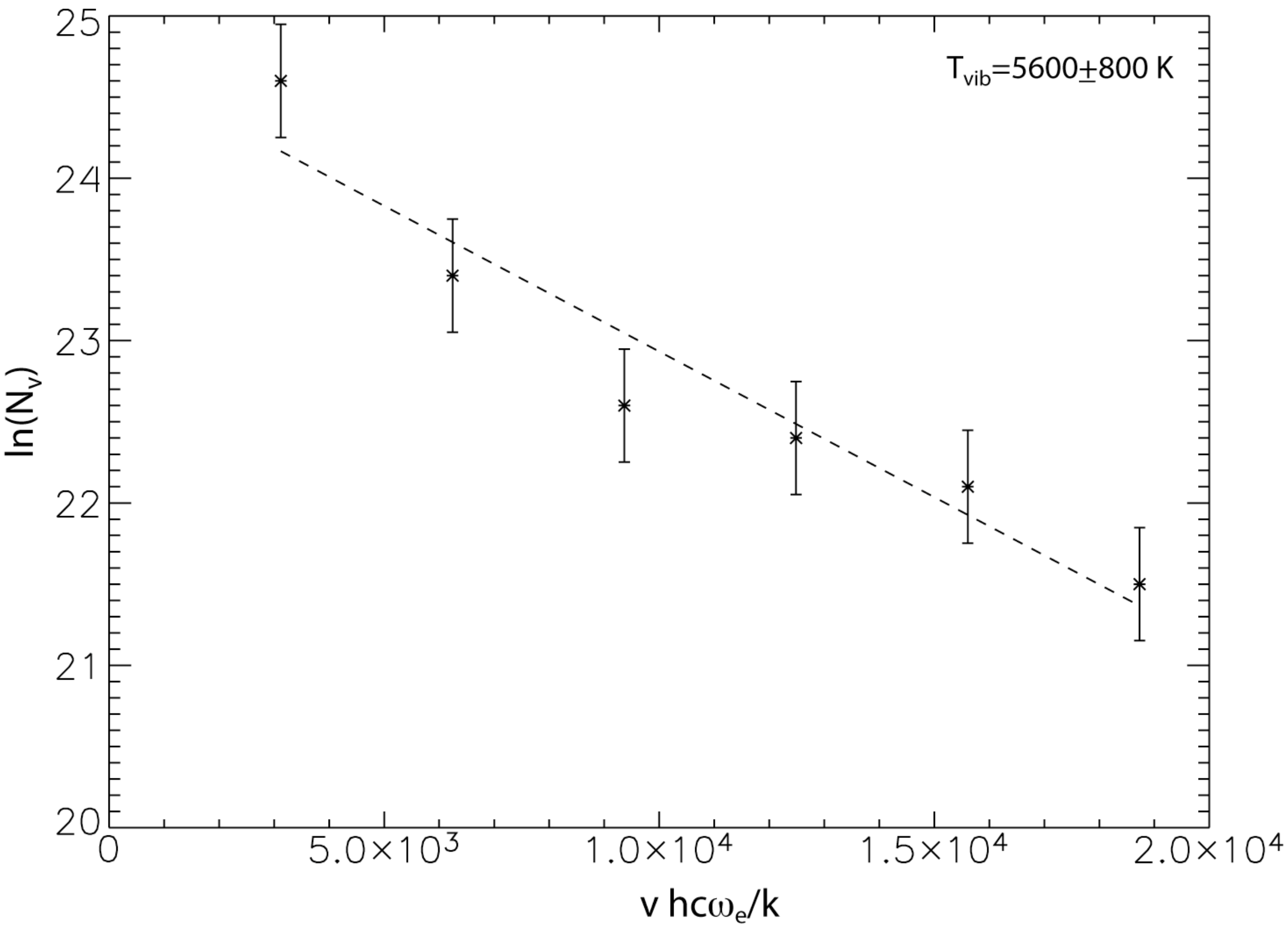

**Figure 8.** Vibrational temperature of $^{12}CO$ for HD 141569. The high vibrational temperature of the gas relative to the rotational temperature (Figure 5) suggests that the gas is excited by UV fluorescence.

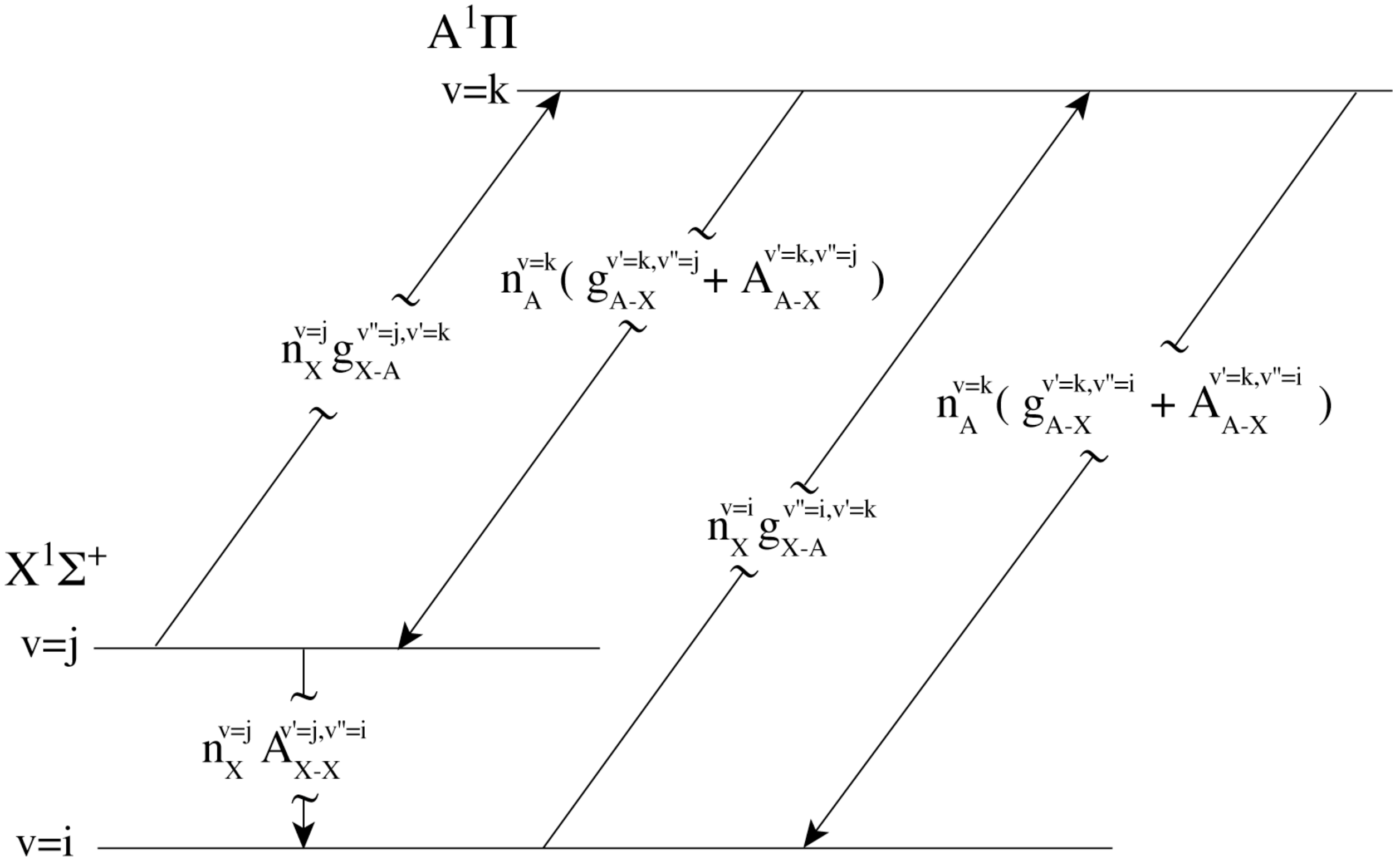

**Figure 9.** Energy level schematic of the CO molecule. In this schematic we highlight the types of transitions we consider in our calculation. The pumping factor, $g_{ij}$, is proportional to the oscillator strength of the transition and the incident radiation field. In our calculation we consider the vibrational levels v=0 through v=9 for the ground electronic state, $X^1\Sigma^+$, and the first excited electronic level, $A^1\Pi$.

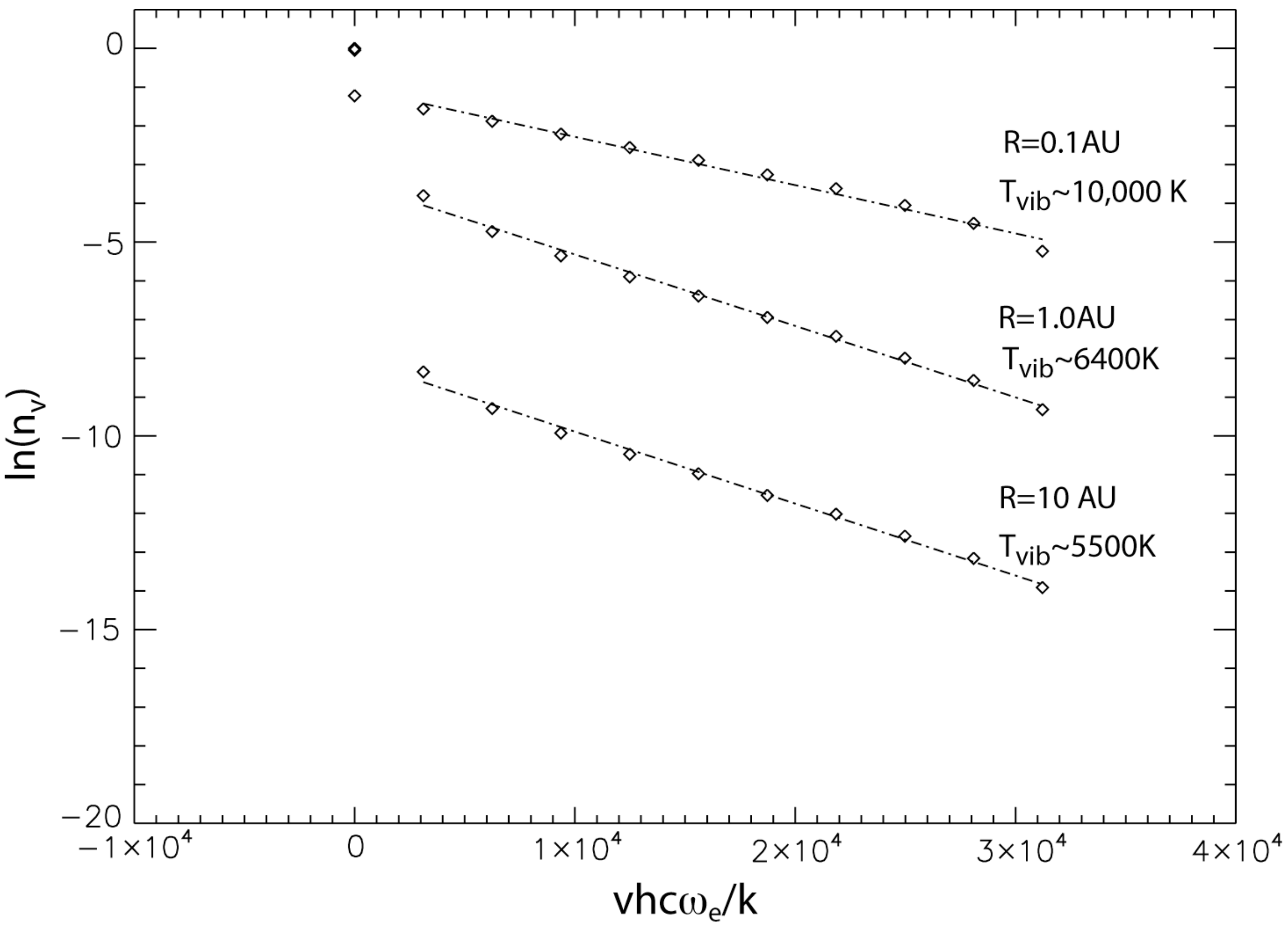

**Figure 10.** The vibrational populations of CO excited by the radiation field of an A0 star at 0.1 AU, 1 AU, and 10 AU. We plot the natural log of the fractional level populations versus the vibrational energy above the ground state such that the vibrational temperature is inversely proportional to the slope of the fit. As the radiation field grows more dilute and the UV pumping is diminished, the ro-vibrational transitions become more effective at "cooling" the vibrational levels of the gas. For comparison, the relative populations of $n_{v=1}$ and $n_{v=2}$ of thermalized CO at 200K are $\ln(n_{v=1})$=-15.6 and $\ln(n_{v=2})$=-31.2.

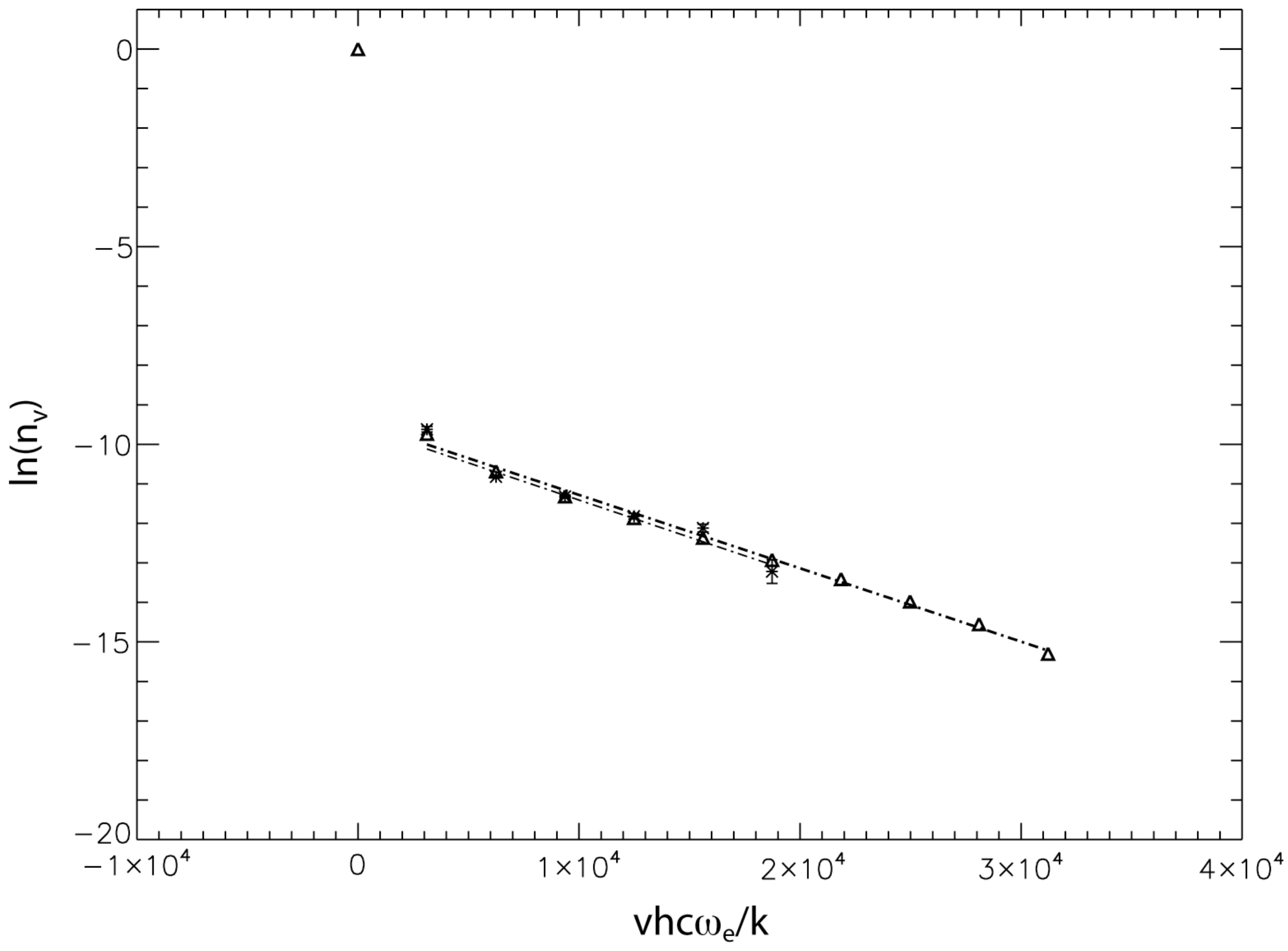

**Figure 11.** Comparison of the calculated fractional vibrational populations to the measured populations of CO around HD 141569. The observed vibrational populations are plotted with asterisks and the calculated levels are plotted using triangles. The measured populations are normalized to the fractional population calculated for v=1 using our UV fluorescence model. The agreement between the prediction and observation confirms that UV fluorescence dominates the excitation of the gas.

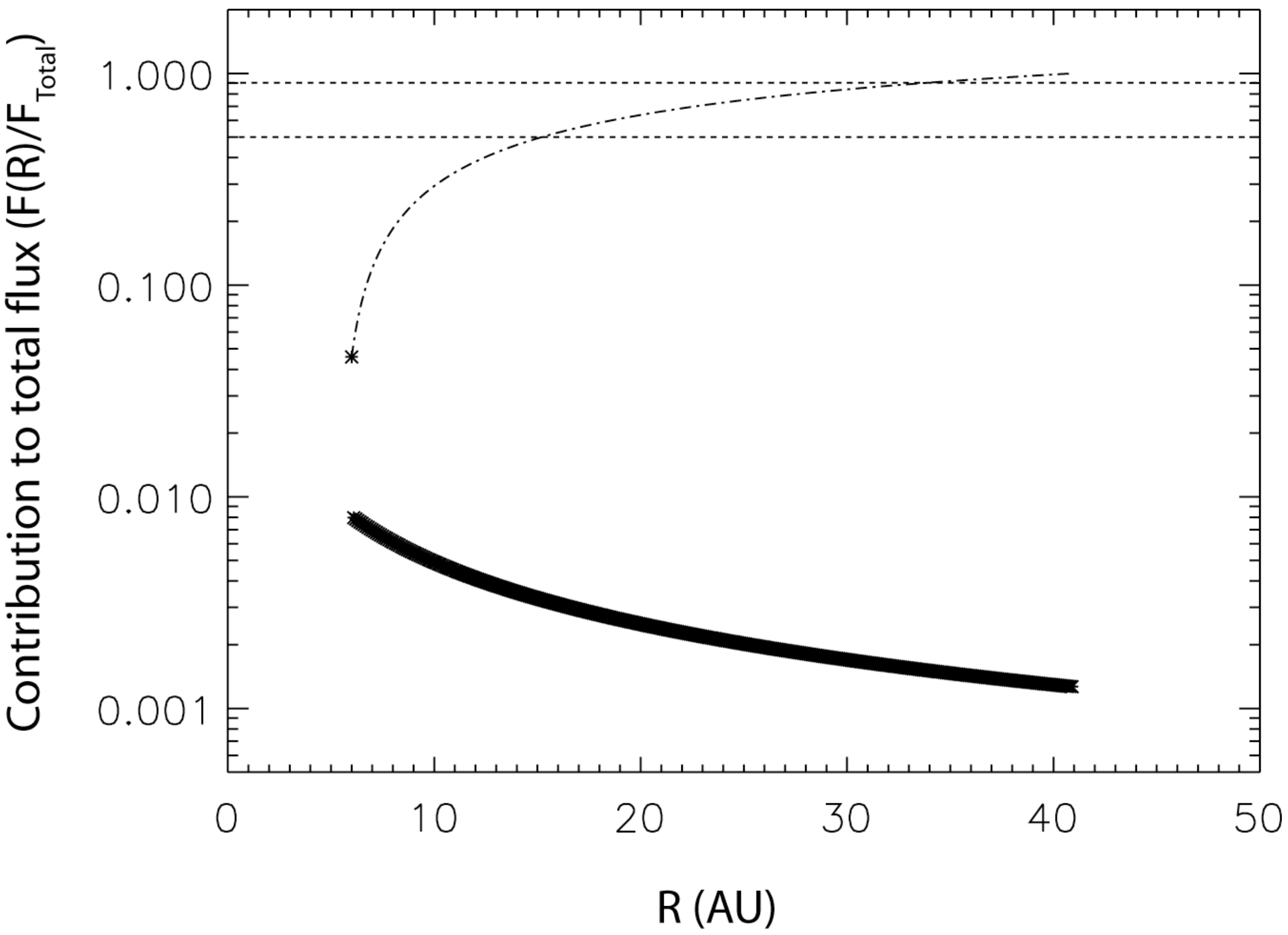

**Figure 12.** Contribution to flux as a function of radius. Here we plot the fractional contribution of each annulus to the total observed flux (asterisks) and the integral of the fractional contribution (dot-dashed line). Half of the flux originates from 6-15AU and 90% of the flux comes from 6-34AU. The 50% and 90% levels are indicated with dotted lines. The inner edge of the disk generates more flux than any other annulus because of its greater exposure to the UV field of the star.

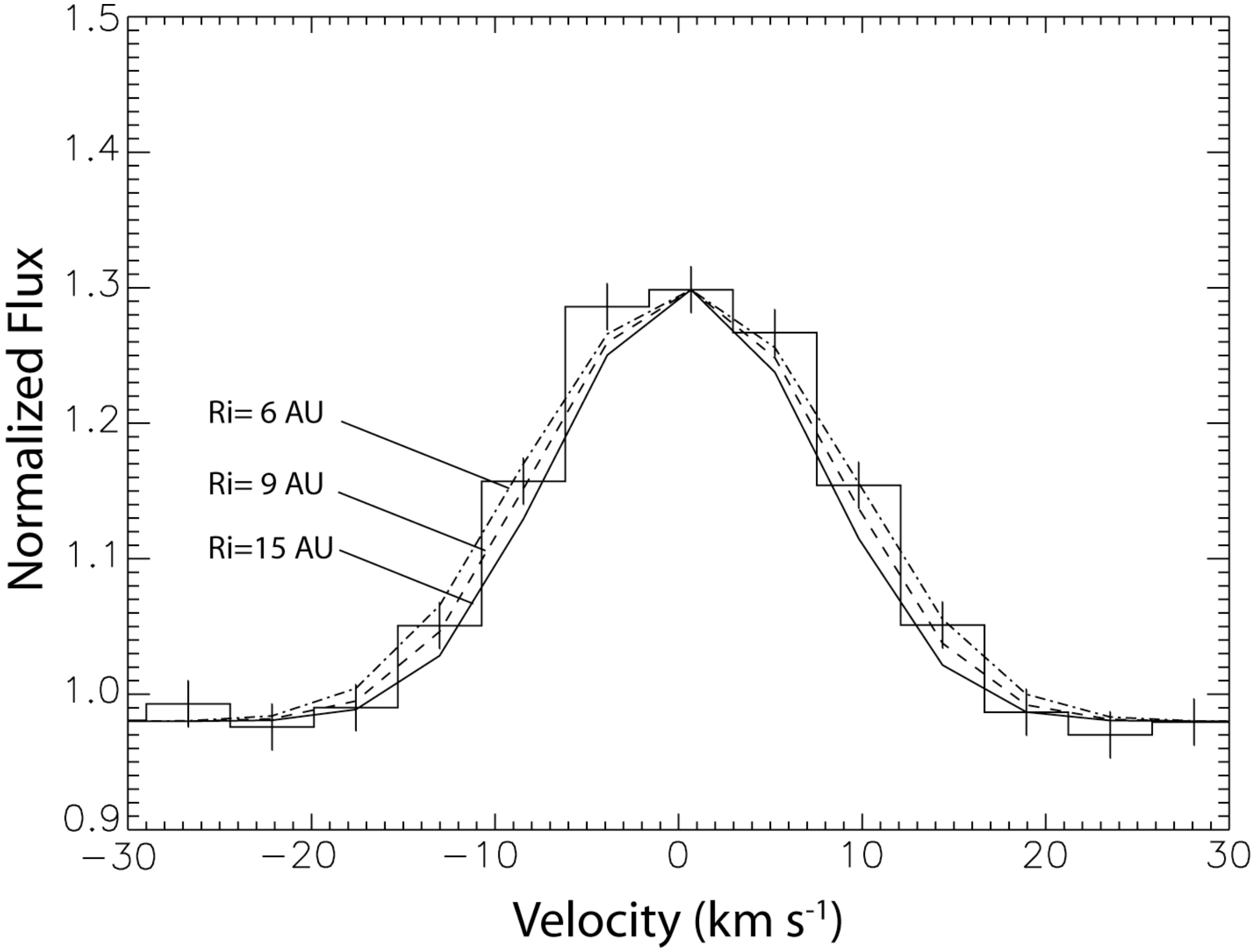

**Figure 13.** Spectral synthesis of the v=2-1 R9 CO emission line. The fits correspond to different inner radii spanning 6AU-15AU. The best fit is 9AU, which is consistent with measurement by Goto et al. (2006) who find the inner rim is 11±2 AU. In our model, we allow the UV flux to penetrate a depth of N(CO)=$10^{15}$ $cm^{-2}$, thus this is our upper limit on the column density of CO interior to 6 AU.

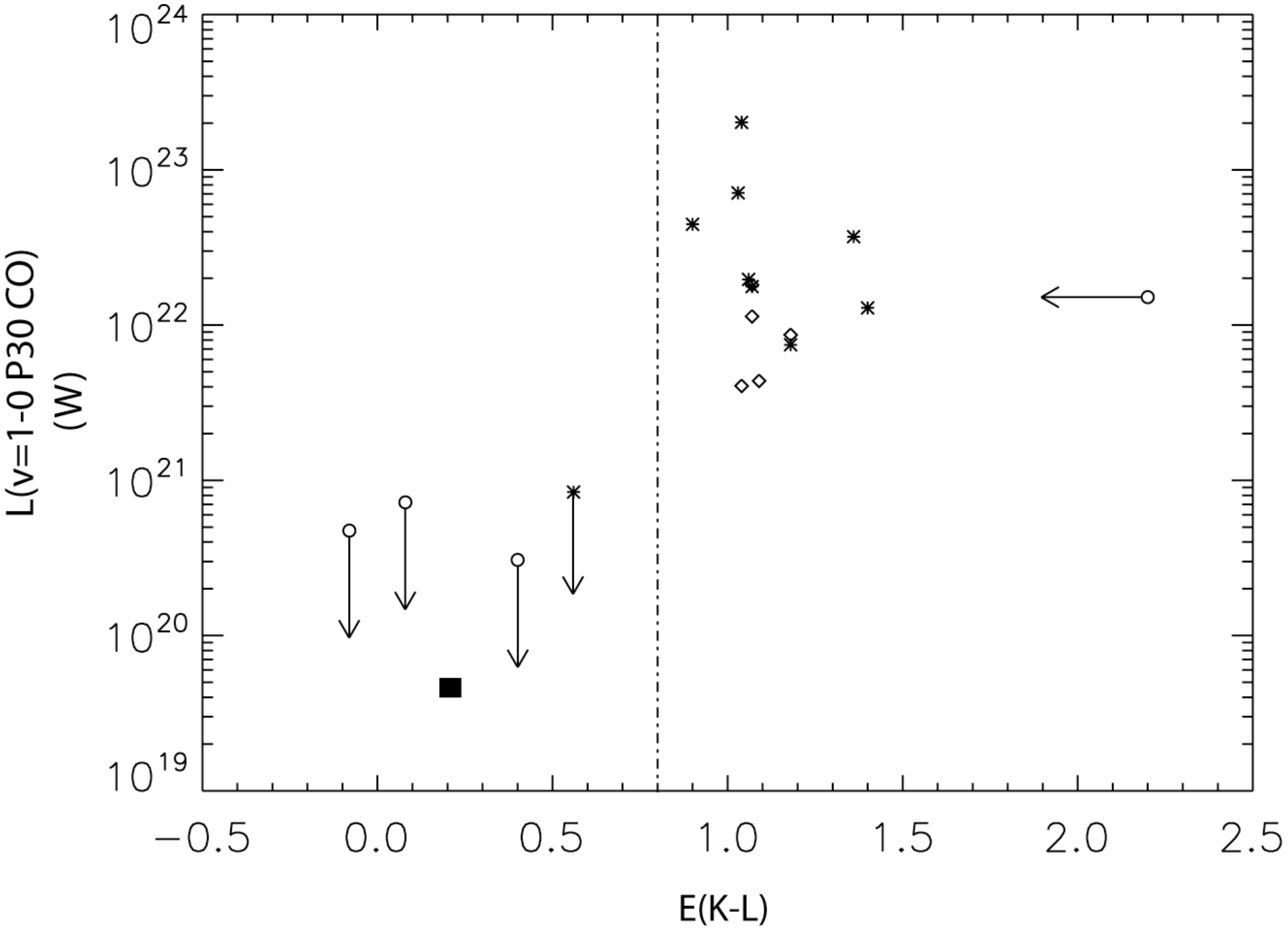

**Figure 14.** CO luminosity versus the NIR excess. All of the sources that reveal a K-L excess of greater than 1 magnitude reveal CO emission. However, only 1 out of 5 sources without hot dust reveals CO emission. The sources for which Br γ is not detected in emission are labeled with open circles. HD 141569, labeled by a box rather than by an asterisk, does not reveal the P30 line in emission. The plotted line luminosity for this source is extrapolated from the observed lines. The NIR excess for Elias 2-22b is presented as an upper limit because its extinction is not known. It is possible that it is a gas rich/dust poor system if the foreground extinction is >30 magnitudes. The diamonds are data points inferred from Blake & Boogert (2004).

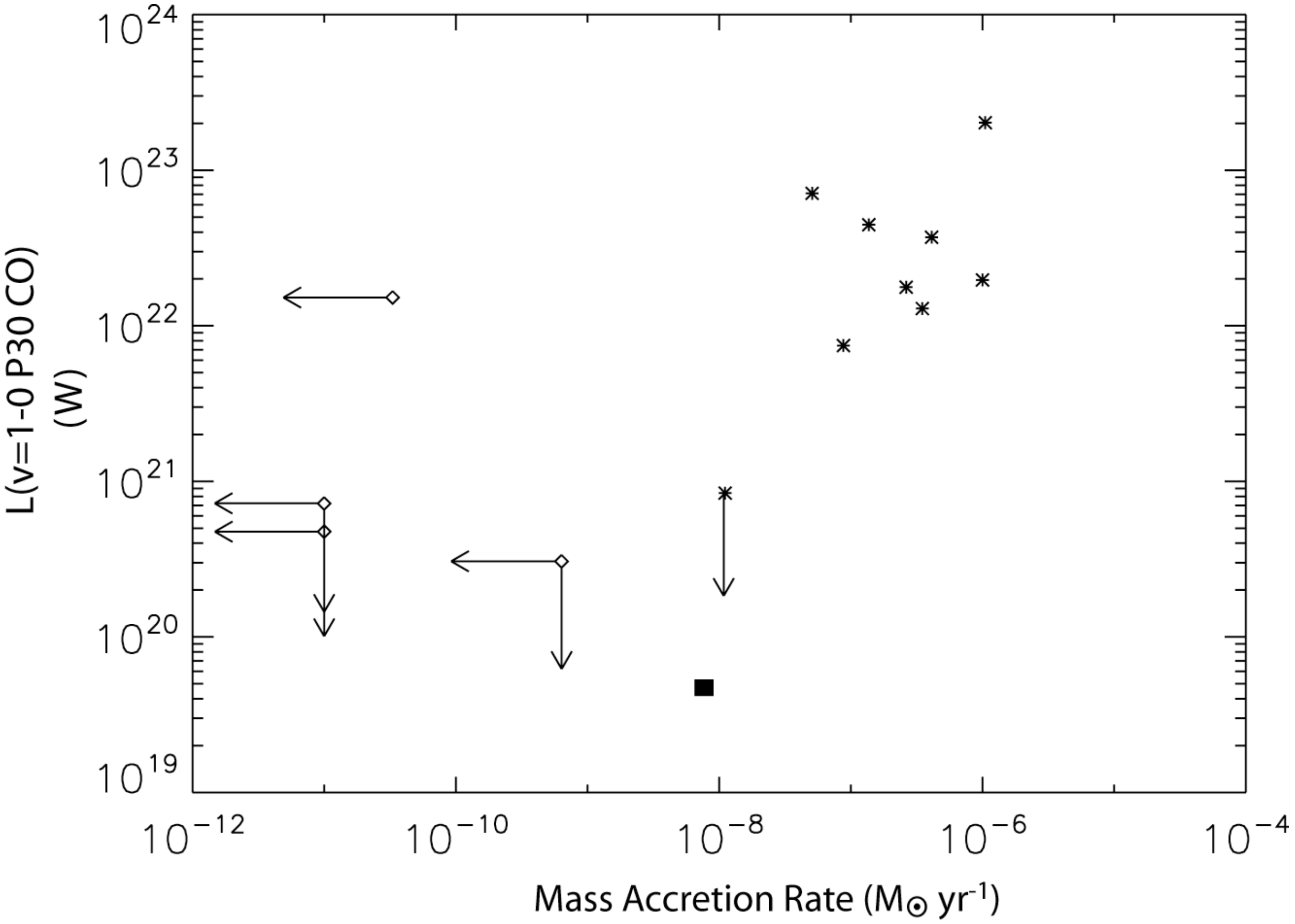

**Figure 15.** CO luminosity versus mass accretion rate. The mass accretion rate for each source was calculated from the Br γ luminosity by applying equations 7 and 8. The luminosities were calculated using the distances presented in Table 3. All of the sources with accretion rates more than a few times $10^{-8}$ $M_{\odot}$ $yr^{-1}$ reveal the v=1-0 P30 CO line in emission. HD 141569, labeled by a box rather than by an asterisk, does not reveal the P30 line in emission. However, lower lying ro-vibrational lines have been observed. The plotted line luminosity for this source is extrapolated from the observed lines. The sources for which Br γ is observed in absorption (HD 38087 and HD 149914) are plotted with upper limits of $\dot{M} \leq 10^{-11} M_{\odot}\, yr^{-1}$. The upper limit on the accretion rate of Elias 2-22b is calculated assuming $A_V$~20.

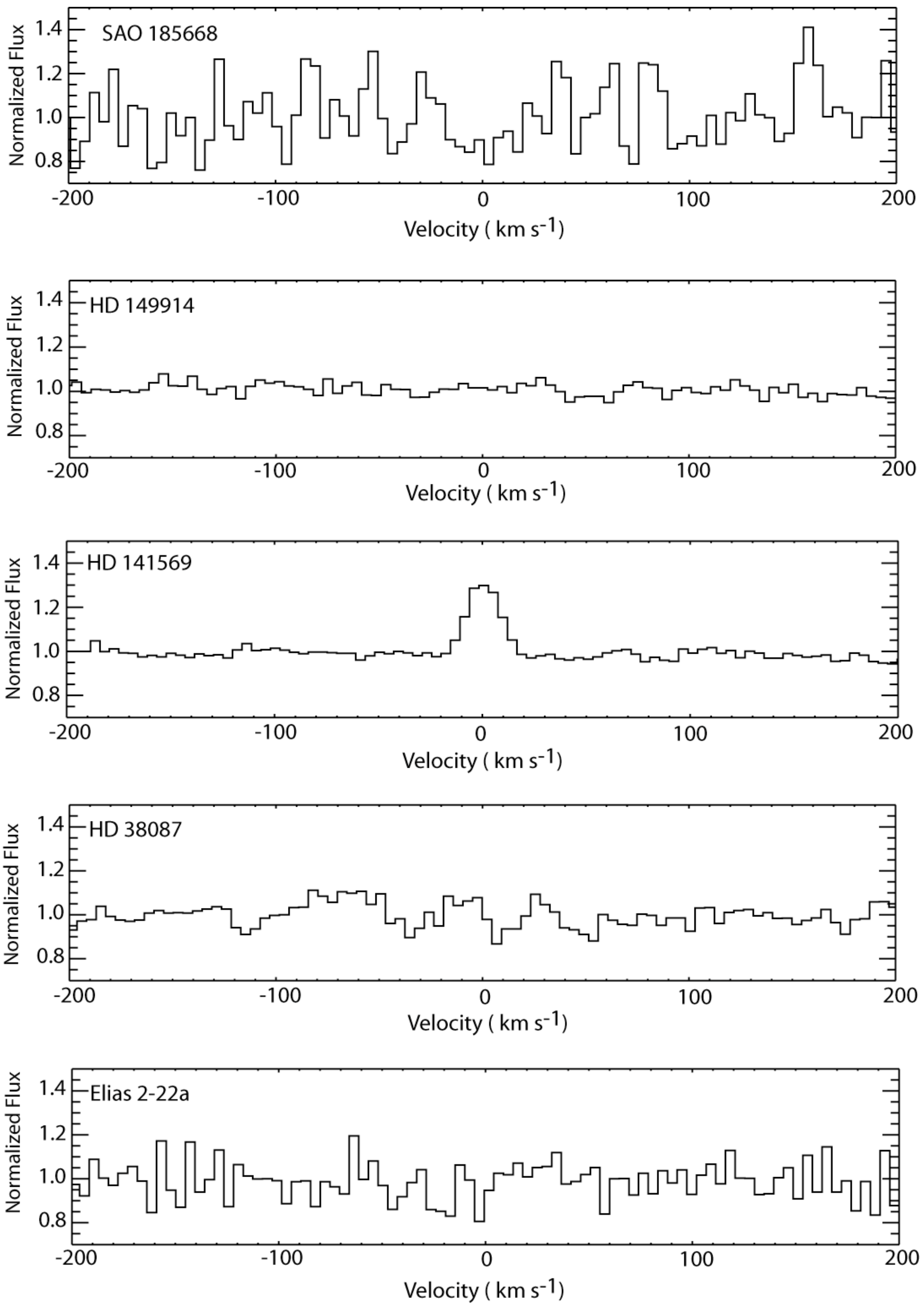

**Figure 16.** Spectra of the R9 v=2-1 line from the five transitional disks. CO emission is only detected in HD 141569.